\documentclass[sn-mathphys,Numbered]{sn-jnl}

\usepackage{graphicx}%
\usepackage{multirow}%
\usepackage{amsmath,amssymb,amsfonts}%
\usepackage{amsthm}%
\usepackage{mathrsfs}%
\usepackage[title]{appendix}%
\usepackage{textcomp}%
\usepackage{manyfoot}%
\usepackage{booktabs}%
\usepackage{algorithm}%
\usepackage{algorithmicx}%
\usepackage{algpseudocode}%
\usepackage{listings}%
\usepackage{pdfpages}
\usepackage{doi}
\usepackage{enumitem}

\begin{document}

\title[QT Industry Needs]{Advancing quantum technology workforce: industry insights into qualification and training needs}

\author*[1]{\fnm{Franziska} \sur{Greinert}}\email{f.greinert@tu-braunschweig.de}

\author[1]{\fnm{Malte S.} \sur{Ubben}}

\author[1]{\fnm{Ismet N.} \sur{Dogan}}

\author[1]{\fnm{Dagmar} \sur{Hilfert-Rüppell}}

\author[1]{\fnm{Rainer} \sur{Müller}}

\affil[1]{\orgdiv{Institut für Fachdidaktik der Naturwissenschaften}, \orgname{Technische Universität Braunschweig}, \orgaddress{\street{Bienroder Weg 82}, \city{Braunschweig}, \postcode{38106}, \country{Germany}}}

\abstract{
The transition of second-generation quantum technologies from a research topic to a topic of industrial relevance has led to a growing number of quantum companies and companies that are exploring quantum technologies. Examples would include a start-up building a quantum key distribution device, a large company working on integrating a quantum sensing core into a product, or a company providing quantum computing consultancy. They all face different challenges and needs in terms of building their quantum workforce and training in quantum concepts, technologies and how to derive value from them. 

With the study documented in this paper, we aim to identify these needs and provide a picture of the industry's requirements in terms of workforce development and (external) training and materials. We discuss, for example, the shortage of engineers and jobs relevant to the quantum industry, the challenge of getting people interested in quantum, and the need for training at different levels and in different formats \textbf{--} from awareness raising and self-learning materials to university courses in quantum systems engineering. The findings are based on 34 semi-structured interviews with industry representatives and a follow-up questionnaire to validate some of the issues raised in the interviews. These results have influenced activities in EU projects, including an update of the European Competence Framework for Quantum Technologies. }

\maketitle

\newcommand{\myquote}[2][]{ \begingroup 
\noindent
\leftskip 20pt \rightskip 20pt 
\small
\textit{#2}
[#1]

\endgroup 
\vspace{0.5\baselineskip}
\noindent
}

\newcommand{\larMulti}{large multi}
\newcommand{\onlyQcomp}{only qComp}
\newcommand{\sme}{SME}
\newcommand{\pureQT}{start-up}

\newcommand{\CF}{CFQT~\cite{greinertEuropeanCompetenceFramework2024}}

\section{Introduction}\label{sec1}
    \myquote[19 \larMulti, Pos. 60\footnote{Quote from one of the interviews analysed in this paper; see Sec.~\ref{subsubsec:interviewParticipants} for reference code explanation.}]{I'm very happy to see that the universities all over Europe are now realising that quantum technologies [...] 
    will be a big game changer. 
    So that [... we as industry] 
    will get trained people. 
    }%
With the growing industrial relevance of modern quantum technologies (QTs), there is an increasing need for a well-educated quantum workforce with specific requirements~\cite{masiowskiQuantumComputingFunding2022}.
At European level, the Quantum Flagship~\cite{qucatsQuantumFlagshipFuture2024} is driving efforts to develop the quantum workforce within the coordination and support action (CSA) QUCATS, for example by  updating and extending the European Competence Framework for Quantum Technologies (CFQT)~\cite{greinertEuropeanCompetenceFramework2024, greinertQuantumReadyWorkforce2023a}. 
A curated playlist\footnote{Direct link to the playlist: \url{https://quantumready.eu/\#/playlist}.} 
of videos has been created as part of the QUCATS project and is available through the European Quantum Readiness Center (EQRC)~\cite{quantumflagshipEuropeanQuantumReadiness}. The EQRC is the central platform for making Europe `quantum ready' by analysis, providing resources and establishing best practices.
Two additional EU-funded projects are the DigiQ project~\cite{shersonDigiQDigitallyEnhanced2024} for developing QT master's programs and the QTIndu project~\cite{qurecaQTInduQuantumTechnologies2024} for developing a program for upskilling industry workforce in QTs. 
More quantum educational efforts are reviewed by Kaur and Venegas-Gomez~\cite{kaurDefiningQuantumWorkforce2022}.
Notably, a number of studies have been carried out in the US to analyze aspects of QT workforce needs at the national level~\cite{foxPreparingQuantumRevolution2020, aielloAchievingQuantumSmart2021,hughesAssessingNeedsQuantum2022,hasanovicQuantumTechnicianSkills2022}.

Initiatives to meet professional training demands require a clear understanding of industry needs. 
In an earlier study, 
we collected competences, requirements and forecasts around the future quantum workforce~\cite{greinertFutureQuantumWorkforce2023}. This 
resulted in the first version of the \CF\ and a collection of predictions around the future industrial relevance and societal and educational impact of QTs. However, it did not take into account, for example, specific training formats or different job roles in QT.

The aim of this study is to enhance understanding of industry needs to facilitate the development of more effective and practical training and education strategies.
As part of QUCATS, we analyze the education and training needs of the European quantum industry by addressing the following \textbf{research questions}:
\newcommand{\ROone}{What challenges does industry face in terms of workforce development \textbf{--} and what strategies are being pursued to attract quantum-educated people?} 
\newcommand{\ROthree}{What QT qualification and training needs are reported by industry? Which training formats are relevant?}
\newcommand{\ROfour}{What training conditions and self-learning formats are preferred by industry? What other needs and suggestions are being discussed by industry?}
\begin{itemize}[leftmargin=1cm]
    \item[RQ 1] \ROone\ (Sec.~\ref{sec:resultsChallenges})
    \item[RQ 2] \ROthree\ (Sec.~\ref{sec:resultsQP})
    \item[RQ 3] \ROfour\ (Sec.~\ref{sec:resultsThree})
\end{itemize}
A condensed summary of the results is available in Sec.~\ref{sec:sum}. 
Furthermore, some results were used to update of the \CF\ to version~2.5 in April 2024, as touched in Sec.~\ref{subsec:qualiProf}. The steps from the initial interview analysis to the final publication are not discussed in this paper as they are beyond its scope.

\section{Methods}\label{sec:methods}
To address these research questions, we decided for a qualitative approach using interviews. 
The interviewees came from a variety of professional backgrounds, such as the CEO or CTO of a start-up, the HR professional or QT expert in a large company, the sales engineer in an SME or the quantum computing consultant. 
This diversity allowed us to capture and evaluate different challenges and needs in the industry. 

\begin{figure}[ht]
    \centering
    \includegraphics[width=85 mm
    ]{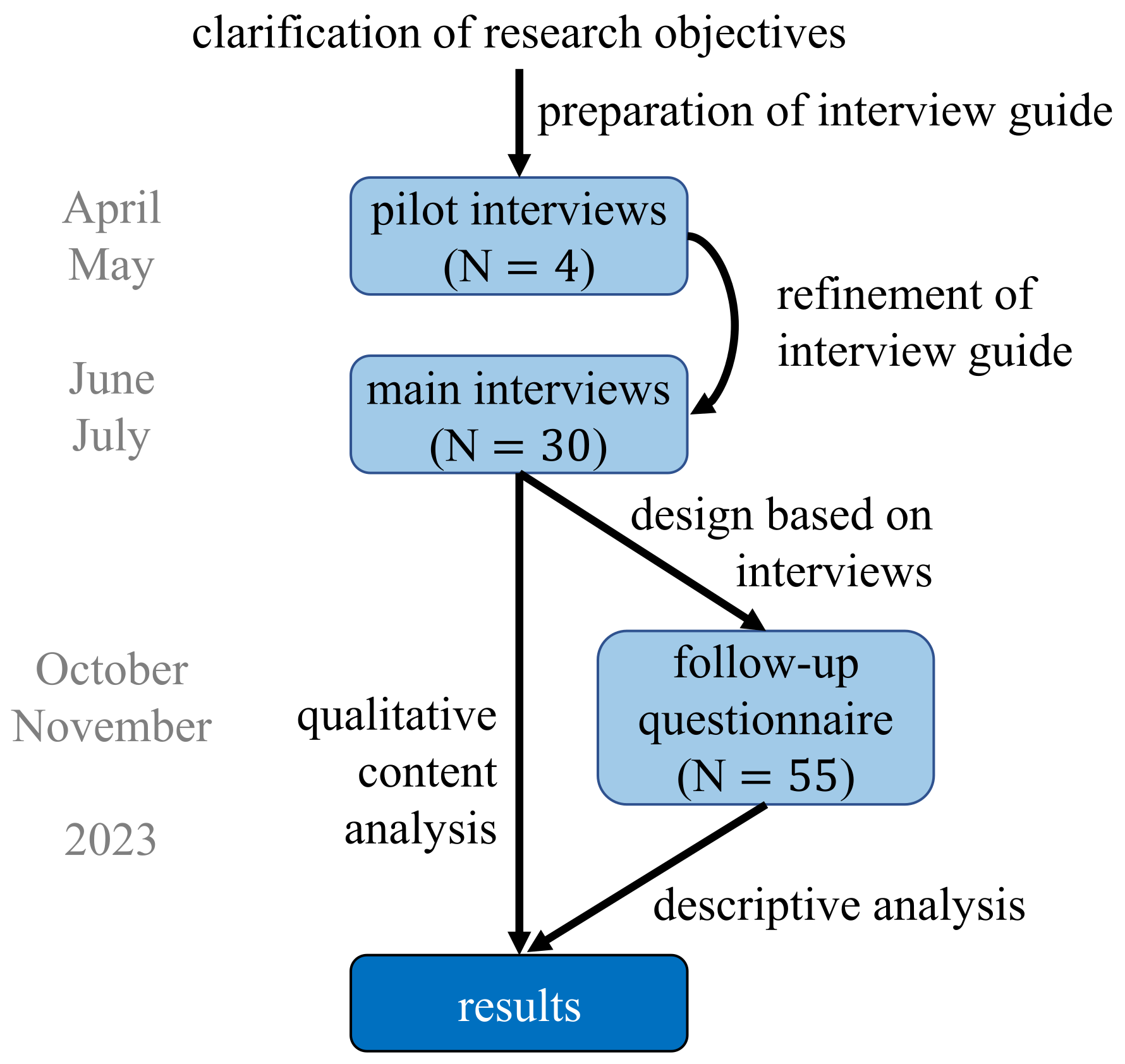}
    \vspace{0.2cm}
    \caption{Study design: Based on the clarification of the objectives, an interview guide was developed and used for four pilot interviews. From this experience, the guide was refined and used for 30 main interviews. An additional follow-up questionnaire was designed based on the results of the initial interviews. 
    Both the main interviews and the 55 responses to the follow-up questionnaire were used to achieve the final results presented in this paper through qualitative content and descriptive analysis.}
    \label{fig:overview}
    \vspace{-0.5cm}
\end{figure}

Some of the topics mentioned by interviewees may be very specific, perhaps even unique to one company. To get a better idea of the generality of some of the needs identified in the interviews, a follow-up survey was created. With this short questionnaire, we aimed at a quantitative validation of the qualitative results, thus following a mixed-methods approach~\cite{cameronSequentialMixedModel2009}. Figure~\ref{fig:overview} provides an overview of the study design.

\subsection{Instruments}
\subsubsection{Interview guide}\label{instrument1}
In preparation for the semi-structured interviews, an interview guide was developed and refined after piloting. The interview guide included the main questions as well as additional maintenance questions. These did not introduce new content but helped to maintain the narrative flow. Furthermore, the guide outlined the expectations of the topics to be discussed for each question. Two slides were prepared to be shown during the interview (Fig.~\ref{fig:CompTypes} and Fig.~\ref{fig:TrainForm}). Most questions and materials were optional and were used depending on the flow of the conversation.
The complete interview guide is available on Zenodo~\cite{greinertSupplementaryMaterialQuantum2024}, an extract is provided as an example in Tab.~\ref{tab:interwGuide}. The main topics are:
\begin{itemize}
    \item Expectations about the role of QT for the own company over the next five years 
    and
    challenges, especially in workforce development (see Tab.~\ref{tab:interwGuide});
    \item skills and job roles relevant in the company in the next five years, might be motivated by displaying material ``competence types'' (Fig.~\ref{fig:CompTypes}) to discuss types and levels relevant for the company in the next five years;
    \item preferred formats for training activities, plus aspects like preferred language (English or local language) and importance of a certificate, might be supported by displaying the material ``training formats'' (Fig.~\ref{fig:TrainForm}).
\end{itemize}

\noindent
Other optional aspects under discussion included the number of new quantum-educated employees expected to be needed and how this demand would be met (new hires vs. upskilling). 
It also covered what degrees might be required and what non-quantum qualifications employees would need to bring. Finally, the interviewees were asked for an advice they would give to educators on how to develop materials or training that are really relevant to the industry.

\begin{table}[hb]
    \centering
    \caption{Extract from the interview guide: first questions to open up the conversation and collect challenges and strategies for RQ~1.}
    \begin{tabular}{p{5.8cm}|p{3cm}|p{3cm}}
    Main question & Additional quest./help & Expectations \\
    \hline
What \textbf{role} do you expect quantum technologies to play in your business over the \textbf{next five years}? [...] 
\newline
What \textbf{challenges} are you currently facing or expect to face in the future in relation to quantum technologies, especially in terms of \textbf{workforce development}? 
&
Which technology or component will you work on? 
[...] 
What specific use cases are you considering? [...] 
&
Statement covering the role (developer, supplier, user, ...) and 
concrete technology/application or use cases that will be in the focus  \\
    \end{tabular}
    \label{tab:interwGuide}
\end{table}

\subsubsection{Questionnaire}\label{instrument2}
The follow-up online questionnaire aimed at validating a number of results from the interviews by involving a larger number of participants with a minimum of time required from them (median processing time 3 minutes). The questionnaire was structured in three blocks:
\begin{itemize}
    \item Background: `Where do you work?' (`industry' or `academia' or `other'; `Europe' or `US and Canada' or `Asia' or `other');
    \item 19 items to assess with `high need' or `low need' or `no need' (or `can't assess');
    \item additional questions about their company for participants from industry.
\end{itemize}
The items are available in Appx.~\ref{subsec:items}.
All questions were in multiple choice format and optional. The survey was implemented via LimeSurvey (\url{https://limesurvey.org/}) and distributed in October/November 2023 by announcing it at the European Quantum Technologies Conference (EQTC), sharing it on LinkedIn and distributing it via established QT networks. 

\subsection{Data analysis}
\subsubsection{Qualitative content analysis of the interviews}\label{sec:ana1}
The interview transcripts were analysed using qualitative content analysis~\cite{mayringQualitativeContentAnalysis2014, radikerFocusedAnalysisQualitative2020} with the software MaxQDA (version 24, \url{https://maxqda.com/}). The category system was developed using a combination of deductive codes, based on the structure of the interview guide (topics discussed) and inductive codes, derived from the responses given in the interviews. It was refined and extended iteratively during the analysis process. The category system for the results (i.e. excluding e.g. the company background categories) is available in Appendix~\ref{app:categories}. The presentation of the results mainly follow the structure of the category system. 

Using MaxQDA, an intercoder agreement of $\kappa = 0.61$ was obtained, a substantial agreement according to Landis and Koch~\cite{landisMeasurementObserverAgreement1977}. The valuation is based on 10\% of the main interviews \textbf{--} selected using an online quantum random number generator \textbf{--} and checking for overlap of the codes of the main coder (who coded all the interviews) with the codes from a second coder. Clarification in the code memos (the rules for assigning specific codes) and subsequent revision by the second coder led to this value of $\kappa$ reported above.

A number of quotes from the interviews are presented in the results sections (Sec.~\ref{sec:resultsChallenges} to~\ref{sec:resultsThree}). 
As these quotes are from recorded oral conversations, the language has been edited to improve readability, for example by correcting grammar and removing filler phrases such as  `you know' and repetition. All emphasis in quotations has been added by the authors.
In the supplementary material on Zenodo~\cite{greinertSupplementaryMaterialQuantum2024} we provide the original (anonymized but not linguistically edited) transcript passages that are reproduced in this paper, as well as the transcript passages  that are referred to in the results sections. In total, they provide 173~transcript passages that form the basis of the results reported in this paper.

\subsubsection{Quantitative analysis of the questionnaire  responses}\label{subsec:DSBC}
A total of 55 responses to the follow-up questionnaire have been collected. They were evaluated using descriptive statistics, specifically Diverging Stacked Bar Charts (DSBC), to illustrate the respondents' opinions. Due to the relatively small number of responses, we decided against further statistical analysis.

DSBC are a special way of presenting data, such as opinions on a rating scale~\cite{robbinsPlottingLikertOther2011}. Each bar contains the number of responses for each opinion option \textbf{--} in our case: `no need', `low need' and `high need' \textbf{--} stacked together to form a single bar. The bars are aligned to the `neutral' opinion, so in our case all bars are aligned to the center of the part representing the responses for `low need'.
A negative number represents the number of people responding to the `negative' half of the scale, thus those people answering `no need' and half of the people answering `low need', as `low need' is the middle of the scale. By comparing the cumulative responses on either side of the scale,  it is visually clear which way the overall opinion is leaning. We have chosen not to normalize the bars and instead present absolute values due to the small sample size. DSBC are used in Fig.~\ref{fig:trainingNeeds}, \ref{fig:langCert} and~\ref{fig:otherNeeds}.

\subsection{Participants}

\subsubsection{Interview participants}\label{subsubsec:interviewParticipants}
Potential participants were contacted in person or by personal message, explaining the aims of the study and requesting voluntary participation. If they agreed to participate, an online interview was recorded. The interviews ranged in length from 15 to 56 minutes, with an average length of 34 minutes. 20 of the N~$=34$ interviews were arranged with the support of the European Quantum Industry Consortium  (QuIC)~\cite{quicEuropeanQuantumIndustry2024}.
Some of the interviews were conducted with two or three people from one company, resulting in 39 participants. 

The participants' companies  were classified into four types (n $=$ number of interviews in this type):
\begin{itemize}[leftmargin=2.5cm]
    \item [start-up] start-ups and scale-ups developing QT \hfill (n $=13$),
    \item [SME] small- or medium-sized enterprise (SME), \newline provider (supplier) of enabling technologies \hfill (n $=5$),
    \item [large multi] (very) large companies with multiple interests in QT \newline (e.g. developing q. sensors, integrating q. communication \newline systems and exploring the use of q. computers) \hfill (n $=7$),
    \item [only qComp] companies interested only in quantum computing \newline  (e.g. use in finance or provide consulting) \hfill (n $=9$).
\end{itemize}
The type, e.g. `\larMulti', is added to the references to the interviews, together with the interview ID (1~to~34) and the position (Pos. X) in the interview transcript. 
Therefore, the `[19~\larMulti, Pos.~60]' at the beginning of the introduction indicates that the quote is from interview~19 with a (very) large company with multiple interests in QT, located at position~60. With this reference, the original transcript passage can be identified in the supplementary material~\cite{greinertSupplementaryMaterialQuantum2024}.

The respondents were located in nine different European countries, most of them in Germany (n~$=13$ interviews), some in France (n~$=6$), UK (n~$=4$), Netherlands (n~$=3$) and Spain (n~$=3$). 
Some of the companies were very large international groups, active in several European and non-European countries. They are counted according to the location of the participants. 

The companies are active in various areas around QT, some of them covering several of the following, thus the numbers sum up to more than 34. Eight interviews each were conducted with companies active in (a) enabling technologies, (b) communication and networks, and (c) sensing and imaging technologies. For quantum computing and simulation, (d) ten companies are active in hardware or software development or in qubit control, (e) five are somehow related to consulting activities and (f) 13 are (interested in) using quantum computing.

\subsubsection{Questionnaire  participants}
Of the 55 questionnaire respondents, 46 were from industry, ten from academia (including four who also work in industry, thus counted in both categories), and three from other fields. They are mainly from Europe (45~responses). 
Only two respondents assigned themselves to Europe and US/Canada, one to Asia, and one to other region of the world. 
Thus, the reported perspective is predominantly European.

\begin{figure}[th]
    \centering
    \includegraphics[width=0.9\linewidth]{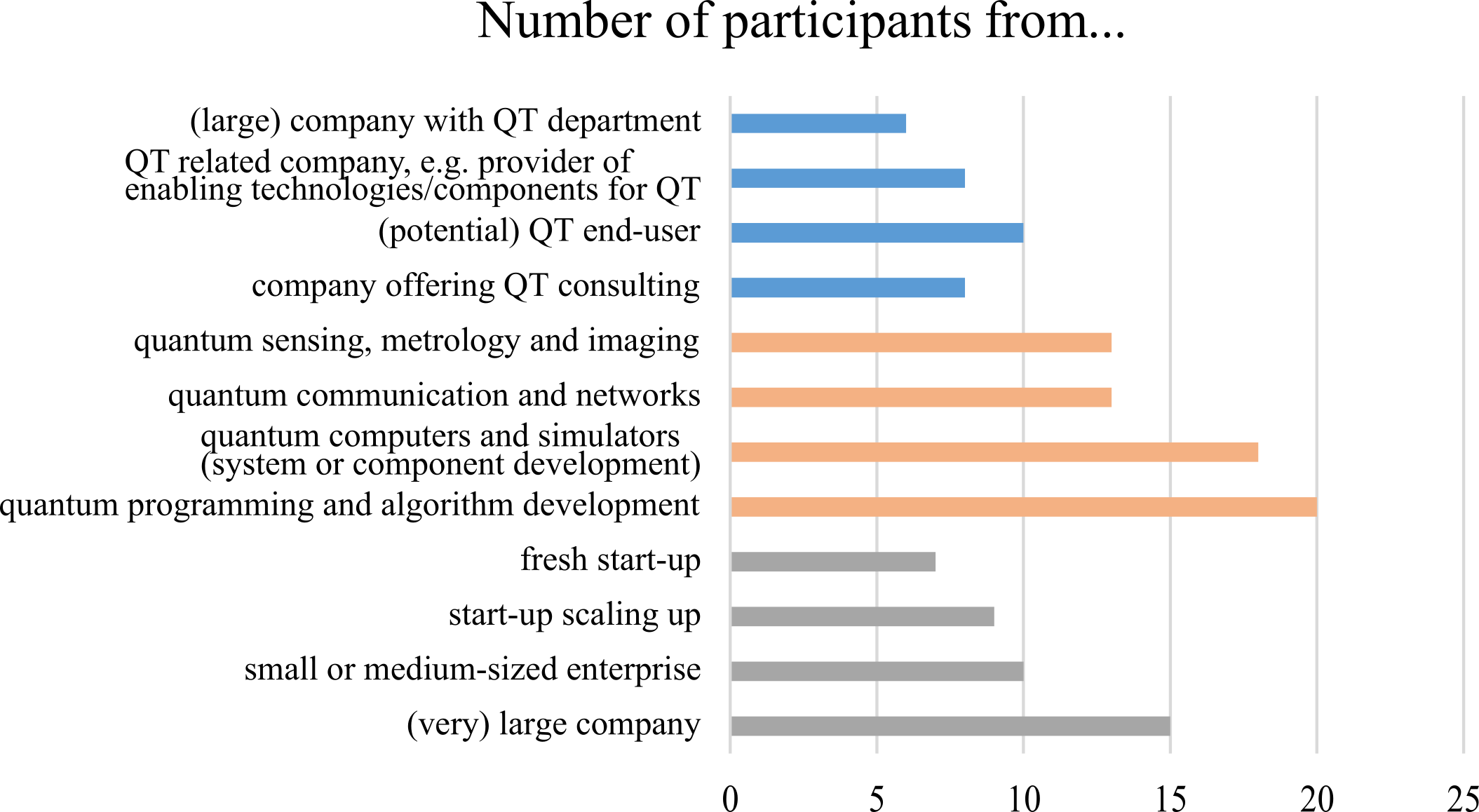}
    \caption{Distribution of the industry background for n~$ = $~41 participants of the follow-up survey that answered this question. Several participants selected more than one of the options in a block, e.g. (very) large companies that cover all four requested QT areas. }
    \label{fig:questPart}
\end{figure}

For the industry participants, some additional background data was requested (optionally). 
Fig.~\ref{fig:questPart} shows these data. The respondents represent a wide range of companies, from start-ups to large corporations, from all QT pillars and active in a variety of functions in the European QT industry.  

\subsection{Limitations}\label{subsect:limitations}
One limitation of our study is related to the interviewees. They cover a heterogeneous group of professions and QT(-related) companies. However, they all come from a small part of the industry: all companies were already active in QT at the time of the interviews. Therefore, it is expected that the responses are biased, as also mentioned by one interviewee [23~\onlyQcomp, Pos.~15]. For companies not yet active in QT, their perspectives would likely differ.

34~interviews is not a small number for a qualitative study, and the heterogeneous approach leads to a diverse collection of opinions. However, we cannot expect to get a complete or representative picture of the QT industry. The group is not evenly distributed, e.g. there were 13~start-ups compared to 5~SMEs. These sub-groups, which may have specific needs, are so small that the validity of looking at them is limited. Similarly, there is a wide range of roles covered by the respondents. To distinguish, for example, between the `manager's perspective' and the `engineer's perspective' would require further analysis with larger groups covering that specific role.

The industry representatives come from Europe, with a strong emphasis on Germany. Such a focus is also visible in other QT-related statistics, e.g. within the Quantum Flagship members (heatmap at~\cite{qucatsQuantumFlagshipFuture2024}). To address this regional focus, we link some of the results to earlier findings from US studies in the discussion (Sec.~\ref{sec:disc}).

Concerning our analysis process, we evaluated the intercoder reliability (substantial agreement, as reported in Sec.~\ref{sec:ana1}). In addition, for the sake of transparency, we provide in the appendix or as supplementary material on Zenodo~\cite{greinertSupplementaryMaterialQuantum2024}:
\begin{itemize}
    \item full interview guide (questions and expectations) (Zenodo);
    \item additional materials for the interviews (Appx.~\ref{app:interwMaterials});
    \item category system for the qualitative content analysis (Appx.~\ref{app:categories});
    \item 173 interview passages (anonymized), covering the quotes and other references to the interviews given in the results (Zenodo):
    \item full assessment items of the follow-up survey with statistical data (Appx.~\ref{subsec:items});
    \item full dataset of the follow-up survey (Zenodo).
\end{itemize}
The follow-up questionnaire validates some aspects identified in the initial interview analysis. Here we combine the analysis of qualitative and quantitative data, following a mixed methods approach~\cite{cameronSequentialMixedModel2009}. 

\section{Challenges \& strategies in workforce development}\label{sec:resultsChallenges}
We start the presentation of the results of the interview analysis with RQ~1: \textit{\ROone}  This section is structured into (1)~challenges in attracting or finding talent,  (2) challenges around QT experts and the need for specific degrees, (3) strategies regarding hiring or upskilling employees, and (4) special challenges for a start-up in scaling up. 
In addition, professions mentioned in the interviews were coded, allowing a quantitative look at the distribution of these about 200 coded segments across a variety of professional groups.
A condensed summary of the presented results can be found in Sec.~\ref{sec:sum}.

\subsection{Talent scarcity}\label{subsec:attractTalent}
\myquote[22~\onlyQcomp, Pos.~16]{The biggest challenge for everyone will be the talent, the talent scarcity which is amazing.}

\myquote[08~\pureQT, Pos.~67]{
The people are out there, but it's hard to get them, 
not because they're not trained or educated enough, it's just that there are not enough of them.}%
Industry representatives from 12 out of 34 companies mention the talent shortage or the difficulty of attracting talent, representing all types of companies. The shortage is typically described as a huge problem [see e.g. quotes above]. However, one of them does not see the shortage as critical yet, but expects it to become more so [6~\larMulti, Pos.~8]. Smaller companies (SMEs or start-ups) would rather like to hire people from the region because people tied to the region stay longer [34~\sme, Pos.~14].
This becomes difficult if they have already recruited from their own network and are therefore facing a shortage, or if large companies are competitors of small companies or start-ups [ibid.]. Others report the challenge of being a less attractive field for talented people to work in [11~\onlyQcomp, Pos. 11; 
04~\onlyQcomp, Pos.~24]. 

However, this challenge of attracting talent does not seem to be specific to QT, there are only a few quantum experts and lots of engineers who are in shortage also outside of QT. One person explicitly reports that it is harder to get traditional engineers to come and work in quantum \textbf{--} while another interviewee sees QT as an argument for attracting talent:

\myquote[12~\pureQT, Pos.~11]{\label{quote:systemArchitect}We are talking always about quantum technologies, but at the end the products are made of standard electronics, digital electronics and analogue electronics, firmware software. All of them are challenging and not just for our segment. 
[...] The \textbf{quantum part is just the system architect}, and for that specific function we have more than enough candidates and already employees that can cover this. 
[...] \textbf{I don't feel that we have a specific case in the quantum technology compared to other technologies.}}%

\myquote[08~\pureQT, Pos.~10]{People that have a quantum physics background or quantum engineering background, they are quite easily to get because they really look for a quantum company. 
[...] 
for us it's mostly \textbf{hard not to get the quantum engineers, but to get to the software engineers}.}%

\myquote[21~\pureQT, Pos.~11]{Quantum control is at the moment 
[...] the most challenging engineering in the world, which is positive 
in terms of attracting talent, because the \textbf{best talent wants to work on the most challenging problems}.}

\subsection{Quantum experts and required degrees}\label{subsec:degrees}
In terms of QT experts, new challenges have arisen recently. For example, experimentalists lack laboratory experience due to COVID and pandemic restrictions [28~\pureQT, Pos.~25]. Furthermore, there is a shortage of postdocs in academia because they are going into industry \textbf{--} while a PhD is becoming less relevant in industry:

\myquote[26~\pureQT, Pos.~77]{
[Academic research group leaders] were struggling to get postdocs because the companies were absorbing them all. [...] 
But it's always nice up to certain level to get trained through \textbf{university because that's where innovation happens}. 
}%

\myquote[26~\pureQT, Pos.~84]{
[...] really need to have PhDs 
stops being the case. 
}

\myquote[20~\larMulti, Pos.~16]{What we need is not a PhD student with a physics background. We need engineers [... having] 
a \textbf{quantum background} or at least some awareness on quantum. 
}%
There is a shift over time, especially in start-ups, from requiring a high level of expertise and experience, i.e. a PhD in a related field, to opening up to masters and the like. As companies or QT departments grow, and as technology readiness increases, there is a growing need for more junior professionals to join the company and develop their careers [e.g.~28~\pureQT, Pos.~22]. 
A roadmap leader needs a PhD and experience, and sales engineers may need a master's in quantum physics or something like that to be able to really talk to customers, which in the early years of a company are mainly research labs with very specific (physical) requirements [e.g. 16~\pureQT, Pos.~8]. Other roles, such as application scientists, may also require a PhD, but the vast majority of people needed are engineers who do not require such a high level of qualification in quantum physics [e.g. 32~\pureQT, Pos.~16 \& 33-34, 26~\pureQT, Pos.~73-76].

A strategy in the SMEs that are providers of enabling technologies is to mix experienced engineers with traditional/classical skills and a few QT experts. For example, they hire fresh PhDs to bring QT expertise into the company, mainly to answer strategic questions such as requirements to enable the use of their product for QT and the broader QT perspective [3~\sme, Pos.~17]. On the other hand, one respondent pointed out that the QT relationship is not yet important, perhaps in ten years' time [25~\sme, Pos.~26]. 
In addition, one photonics company reported that their quantum people have moved to a QT company, so they have to rely on networks to get information about the relevance of QT to their company [34~\sme, Pos.~8 \&~35]. 

\subsection{Scaling-up the start-up}\label{subsec:startScale}
\myquote[32~\pureQT, Pos.~13]{\label{quote:logistics}%
At the beginning with any start-up, it's heavily R\&D focused. 
Then you 
need to attract [...] 
employees that really know the nitty gritty details. But as the company grows and 
scales, you need 
[... a more] diverse workforce with 
different disciplines, 
logistics, operations, finance, HR. [...] 
with the hiring of 
more generalized roles, the knowledge gap between what their profession is versus what the company does as a whole is increasing. So we also see a need to \textbf{make quantum more easily accessible} for people that do not have a quantum background.}%
Several QT start-ups report strong relationships with universities, some are university spin-offs. At the beginning they have a high academic background, have experimental physicists with many years of laboratory experience and experienced engineers who bring the required expertise. In the beginning, everyone does (almost) everything, and it is a bonus if, for example, there is a financial expert who takes care of all the aspects that the physicists or engineers are not experienced in, so that they can concentrate on their main work [26~\pureQT, Pos.~37]. 

At this early stage, the main challenge is to transform a university or research lab system into an industrial product, e.g. by miniaturizing and scaling it, adding a user interface, or making it run reliably 24/7 at the push of a button [29~\pureQT, Pos.~37]. 

The next challenge is to make industry aware of the existence of the product (and the whole company), especially small university spin-offs miss these contacts, the first customers are usually academic or research labs [16~\pureQT, Pos.~6].

As the company grows, it needs not only more technical staff for production, but also more operational and administrative staff.  So there is a need to make quantum more easily accessible, also to the non-technical people [32~\pureQT, Pos.~13, above]. 
At this stage, some companies start intensive recruitment programs [18~\pureQT, Pos.~6] and open up to people from different backgrounds as they begin to have the capacity to train [28~\pureQT, Pos.~23]. The challenge is the widening knowledge gap between the experts and the new recruits. While an intermediate, e.g. an engineer with some QT experience can follow and learn from an expert, a novice will not be able to do so, and the expert will not be able to train the novice (due to capacity and perhaps lack of skills to explain and train) [32~\pureQT, Pos.~19/quote on p.~\pageref{quote:advancedExpertsGap}]. 
In addition, the intermediate level of engineers and the like is hard to find:

\myquote[10~\pureQT, Pos.~15]{\label{quote:multiChemistry}
It's definitely hard to find people with the right combinations of skills that we need. 
[...] 
Getting engineers that already know about quantum technologies [... is challenging as there are] 
less [...] 
than physicists. 
Having \textbf{engineers or chemists that are aware of quantum technologies, the opportunities and impact}
, and even knowing some of the nomenclature, the definitions \textbf{--} that would be great. 
}%

\subsection{Strategies to meet industry needs: Recruitment, training and  awareness raising}\label{subsec:hireUpskill}

\myquote[20~\larMulti, Pos.~26]{\label{quote:systemEngineer}We have engineers, a lot of engineers today, we have systems engineers, we have IT, we have electric [...] 
working on [...] 
technologies today that are very advanced. 
[...] we are thinking about [...] 
a long term program of education [... for] 
engineers. We don't want to hire all the people from outside [... but also] 
\textbf{upskill people inside the company} that are not at all today working on quantum
.}%
Most companies report a mixed strategy of recruitment and training. Start-ups are a special case, as they don't have the capacity for upskilling, but this may change once they start to scale up [28~\pureQT, Pos.~23]. A lot of upskilling is currently done on the job. One reason is that the company's needs are too specific for external training due to the lack of standardization and normalization of QTs and their components [13~\larMulti, Pos.~36-38].
Some companies are planning long-term upskilling programs, including a `train the trainers' strategy, where future trainers for internal training are trained at universities [20~\larMulti, Pos.~26 (above) \& 29]. 

A difficulty with upskilling programs is that certificates or similar qualification records may not be recognized in the hiring process if they are not considered credible [13~\larMulti, Pos.~57-63], see also Sec.~\ref{subsec:certificate}. 

Companies that are just exploring the use of quantum computing report an unclear roadmap for workforce development [e.g. 5~\onlyQcomp, Pos.~8]. There is uncertainty about business growth, and they lack people with some years of experience in quantum and classical programming, people who are not the high-level experts but have enough expertise for supervision and consulting [22~\onlyQcomp, Pos.~20]. One strategy is to start reskilling classical people now, as it takes time anyway to get the different mindset for quantum:

\myquote[11~\onlyQcomp, Pos.~8]{
I think it's definitely necessary to upskill the people now. 
The quantum topic is very different from doing traditional computing, so the people must be prepared. They need time to learn 
[...] at the moment we see many hybrid methods on the market, hybrid between classical methods and quantum methods. And \textbf{to use this hybrid methods you have to know about quantum computing}.%
}%
This upskilling will build up a hybrid (quantum-classical) expertise, keeping people engaged and motivated in quantum computing while they continue their `classical' careers [31~\onlyQcomp, Pos.~13]. However, there are also doubts on the possibility of upskilling, as the quantum mindset may be too different [22~\onlyQcomp, Pos.~19/quote on p.~\pageref{quote:differentMindset}]. One participant notes that there will not be much investment until the value is visible: primarily the very large companies have the capacity to start with quantum computing already, but for SMEs it will take until the real value can be shown [23~\onlyQcomp, Pos.~12].

Some companies need highly specialized people in a niche field. Instead of QT training, they are more interested in ensuring that universities `deliver continuously' enough people [13~\larMulti, Pos.~7]. 
However, even large companies mainly need engineers (not physics PhDs) with at least some QT awareness [e.g. 20~\larMulti, Pos.~16, previous Sec.~\ref{subsec:degrees}].

Start-ups as well as large companies use relationship to academia to get students for master projects, internships, PhD programs and if they are satisfied try to hire them afterwards [e.g. 17~\pureQT, Pos.~11; 28~\pureQT, Pos.~30; 13~\larMulti, Pos.~34]. QT companies have to search internationally due to talent scarcity [e.g.~28~\pureQT, Pos.~4]. 
Competitions are a strategy for attracting and identifying talent, used mainly by very large and well-known companies. These companies announce competitions in which external talents have to find solutions to industrial research problems. This allows the company to discover new talents. [13~\larMulti, Pos.~73] 

An alternative to hiring own experts is to partner with external experts, such as those in legal or financial fields [28~\pureQT, Pos.~46]. 
In contrast, end users of quantum computing, for example, who are too small to have their own experts, will rely on consultants or service providers [11~\onlyQcomp, Pos.~16].

Decision makers are another relevant group to potentially receive training. In start-ups, they are usually experts in everything the company does [e.g. 26~\pureQT, Pos.~4]. However, as decision makers in very large companies or politicians cannot be experts in all areas, they will ask their experts about QT and rely on expert advice [02~\onlyQcomp, Pos.~43; 31~\onlyQcomp, Pos. 37-39]. Nevertheless, it is important that they see the relevance of QT for their business in order to prepare the company for QT and to provide resources to the employees in QT, e.g. time to focus on it, as well as understanding the overall situations, national capabilities, or e.g. the risks from export control [04~\onlyQcomp, Pos.~72; 30~\larMulti, Pos.~24], as well as the impact:

\myquote[17~\pureQT
, Pos.~34]{The \textbf{decisions makers have to be aware} that we are living in a disruptive era, [...] what we are going to face is just like an industrialization change. [...] Quantum era is going to change the whole lifestyles in the whole world. [...] They have to be adaptive to this change in order to survive.}%

\subsection{Professions discussed in interviews}
In addition to the qualitative analysis, the coding of professions discussed in the interviews allow a quantitative analysis.
Table~\ref{tab:professions} provides this statistic. A total of 288 codes were given for these professions. For the analysis, only one coding of each category per interview was counted to indicate the number of interviews in which the profession was discussed. This results in a total of 196 codes shown in the table.
\begin{table}[ht]
    \centering
    \caption{This table provides the number of interviews with coded segments for the professions (subcategories for category 1.2, see Appx.~\ref{app:categories}), along with the proportion [0.X], rounded to the nearest tenth instead of using percentages due to the small group sizes (ranging from a minimum of 5 to a maximum of 13 interviews per group). \newline 
    (STEM: Science, Technology, Engineering and Mathematics) }
    \begin{tabular}{lcccc}
        &\pureQT &\sme&\larMulti &\onlyQcomp \\
       & (n~$=13$)   & (n~$=5$)   & (n~$=7$) & (n~$=9$)\\
    \hline
        \textbf{engineering}&&&&\\
            electronics, electrical engineering&8 [0.6] &2 [0.4]&3 [0.4]&1 [0.1]\\
            mechanics&4 [0.3]&1 [0.2]&1 [0.1]&
            \\
            `quantum' engineer (explicitly named)&2 [0.2]&1 [0.2]&
            \\
            software, computer science, IT&12 [0.9]\hphantom{1}&2 [0.4]&3 [0.4]&6 [0.7]\\ \quad    `quantum' SW eng. (explicitly named)&2 [0.2]&1 [0.2]&
            &3 [0.3]\\
            systems engineer&3 [0.2]&
            &3 [0.4]&
            \\  
            technical people, fabrication&3 [0.2]&
            &1 [0.1]&1 [0.1]\\
            engineer not further specified&5 [0.4]&3 [0.6]&4 [0.6]&3 [0.3]\\
            other engineers&5 [0.4]&
            &2 [0.3]&
            \\
            \quad    optics/photonics&
            &
            &2 [0.3]&
            \\
            \quad    microwave&2 [0.2]&
            &1 [0.1]&
            \\
        \hline
        \textbf{physics}&&&&\\
            physics, not further specified&2 [0.2]&2 [0.4]&2 [0.3]&6 [0.7]\\
            experimental physicists, lab, hardware dev.&5 [0.4]&
            \\
            `quantum' physicist (explicitly named)&8 [0.6]&
            &1 [0.1]&
            \\
            other physics&2 [0.2]&2 [0.4]&2 [0.3]&1 [0.1]\\
        \hline
        \textbf{other STEM,}&&&&\\         
         experts in the industry domain&
         &2 [0.4]&2 [0.3]&4 [0.4]\\
        \quad    `expert'&1 [0.1]&1 [0.2]&2 [0.3]&
        \\
        \quad    chemistry&3 [0.2]&1 [0.2]&
        &2 [0.2]\\
        \quad    mathematics, also financial&
        &1 [0.2]&
        &3 [0.3]\\
        project/product manager&2 [0.2]&
        &2 [0.3]&
        \\
        sales, marketing, customer engineers&7 [0.5]&2 [0.4]&3 [0.4]&2 [0.2]\\
        \hline
        \textbf{business-related}&&&&\\
            business and (people) management, HR&7 [0.5]&2 [0.4]&2 [0.3]&4 [0.4]\\
            logistics, finance, laws, etc.&6 [0.5]&
            \\
        \hline
        other roles and jobs&5 [0.4]&1 [0.2]&2 [0.3]&2 [0.2]\\
%
    \end{tabular}
    \label{tab:professions}
\end{table}%
About half of all codes (48\% or 95 out of 196) are distributed over the 13 interviews with start-ups (13 out of 34, thus 38\%). For these start-ups, the average number of codes per document is significantly higher, at approximately 7, compared to 4 to 5 in other company types. Thus, start-ups speak more about different roles than other companies. 

Overall, almost half of all roles coded belong to engineering ($46\%
$), less than every fifth to physics ($17\%
$) and only $11\%\,
$ address business roles. 
By far, the most frequent category is the one of \textbf{software engineers}, including data scientists, algorithm designers, programmers, and computer scientists. These professions are needed for tasks such as automation and machine learning or, for example, coding for FPGAs (Field-Programmable Gate Array) [e.g. 29~\pureQT, Pos.~37; 28~\pureQT, Pos.~16; 18~\pureQT, Pos.~16]. 
Only a few companies explicitly mention quantum computer scientists or quantum algorithm developers, e.g. at the error correction level [5~\onlyQcomp, Pos.~28; 21~\pureQT, Pos.~18], 
or explicitly state that they are needed less (in terms of numbers)  than `classical' software engineers [14~\sme, Pos.~10]. However, the `classical' software engineers may benefit from understanding some underlying physics, e.g. for code debugging:

\myquote[16~\pureQT, Pos.~15]{
[It is] sometimes very complicated 
to check the physics behind the code if everything is right or not [...]. 
We need people who understand really good programming and a little bit physics behind this [so] they can prove the code.
}%
A variety of \textbf{engineering professions} are discussed in the interviews. Electrical and electronic engineers, including electromagnetic wave engineers or engineers for micro- or analogue electronics, e.g. for qubit control, are mentioned more often than e.g. mechanical engineers, microwave engineers, optical engineers, or `other engineers' like FPGA engineers, cryo or telecom engineers [e.g. 6~\larMulti, Pos~15; 8~\pureQT, Pos.~10; 18~\pureQT, Pos.~12; 27~\larMulti, Pos.~14]. 
Also, technical people, e.g. technical constructors, are addressed [17~\pureQT, Pos.~61]. Explicit mention of `quantum' engineers is very rare [e.g. ibid], while it is often not specified what kind of engineers are meant when talking about engineers [e.g. 5~\onlyQcomp, Pos.~28].

(Quantum) \textbf{systems} and application engineers, success engineers or system architects are discussed by start-ups and large companies who also work on QT development and integration [32~\pureQT, Pos. 16; 12~\pureQT, Pos.~11/quote on p.~\pageref{quote:systemArchitect}; 20~\larMulti, Pos.~26/quote on p.~\pageref{quote:systemEngineer}; 30~\larMulti, Pos.~9/quote on p.~\pageref{quote:QTsystemEng}]. 
Regarding \textbf{physicists}, experimental physicists and quantum physicists, e.g. on quantum optics or quantum theory, are mentioned [e.g. 18~\pureQT, Pos.~12; 26~\pureQT, Pos.~4; 28~\pureQT, Pos.~25; 29~\pureQT, Pos.~37]. `Other' physicists discussed are, for example, theoretical physicists and cryogenic physicists [29~\pureQT, Pos.~37; 13~\larMulti, Pos.~7]; also physicists are not further specified several times [e.g. 2~\onlyQcomp, Pos.~28; 5~\onlyQcomp, Pos.~28].
Other mentioned STEM roles cover disciplines like  mathematics and finance or also biology and chemistry/material science as application field for QT [e.g.~2~\onlyQcomp, Pos.~28, 5~\onlyQcomp, Pos.~28]. In addition, not further specified (advanced) experts, research scientists or integrators are mentioned [e.g. 32~\pureQT, Pos.~19/quote on p.~\pageref{quote:advancedExpertsGap}].

In start-ups, project/product managers or similar may be physicists now learning management [32~\pureQT, Pos.~44].
In addition, \textbf{sales and marketing} related jobs are mentioned several times. These roles include sales manager, (quantum/technical/direct) sales engineer, PR (Public Relations) and marketing, application scientist, customer engineer or customer interaction scientist, many of whom appear to be (former) engineers who have moved into such customer-oriented roles, and some of which require a very deep understanding of the technologies, while others do not [e.g. 02~\onlyQcomp, Pos.~47/quote on p.~\pageref{quote:salesAsk} 20~\larMulti, Pos.~41; 28~\pureQT; Pos.~46; 29~\pureQT, Pos.~16; 32~\pureQT, Pos.~27 \& 33-34; 33~\pureQT, Pos.~48].

Business and people \textbf{management}, CEOs and decision makers or HR are mentioned by all company types [e.g. 06~\larMulti, Pos.~10; 30~\larMulti, Pos.~24], start-ups discuss also roles in e.g. logistics, finance and legal [e.g. 28~\pureQT, Pos.~46; 32~\pureQT, Pos.~13/quote on p.~\pageref{quote:logistics}].

Other roles mentioned are business case developers, operations and manufacturing or people who oversee production and develop applications [e.g. 07~\onlyQcomp, Pos.~20/quote on p.~\pageref{quote:overseeProduction}]. Furthermore, a quantum analytics translator [23~\onlyQcomp, Pos.~37/quote on p.~\pageref{quote:analyTransl}], and consultants to work with and better understand demands [14~\sme, Pos. 13] are discussed. In addition, start-ups mention e.g.  lab managers [28~\pureQT, Pos. 26] and test engineers who test the equipment before it is shipped to the customer [08~\pureQT, Pos. 14].

\section{Training formats and needs}\label{sec:resultsQP}
This section addresses RQ~2: \textit{\ROthree} The results  indicate that there are  three fundamental needs: basic awareness, targeted training and study programs. 
In addition, general conditions for training, such as the language of the training and the relevance of a certificate, and finally self-learning formats are discussed. The results of both the interviews and the follow-up survey are included in this section. It concludes with an outlook on the \CF\  update with qualification profiles. 

\subsection{QT awareness, impact and hype}
\myquote[17~\pureQT, Pos.~67-69]{Raise the awareness [...] also to the common people [...]. We are facing a change of an era and we really need to do something. I think it's going to be as important also as climate change. The impact. [...] We just are like trying to understand the consequences of the change. [...] Besides dissemination, I think it's important also to raise a very strong \textbf{awareness about the ethics and power of these technologies}, because I fear also, if this is not in the right hands or in the ethical hands, what could happen?}%
The rise of QT, and quantum computing in particular, is expected impact not only industry and academia/science, but also society as a whole.\footnote{
This is an aspect that we have already addressed in an earlier questionnaire-based study, documented in Ref.~\cite[][Sec.~IV.C.4.
]{greinertFutureQuantumWorkforce2023}. Here, participants were asked to assess statements such as \textit{QTs will lead to social inequality}. For this statement, the opinions were very heterogeneous, with a slight majority of `rather agree'.} Although we did not ask about society in general in the interviews, the above comment is an addition made by one interviewee at the very end of the interview. 
It emphasizes the importance of involving and educating society as a whole, creating a basic QT awareness, and considering ethical aspects. However, basic awareness of the expected impact of QTs is particularly relevant for decision makers, as discussed at the end of Sec.~\ref{subsec:hireUpskill}.

With the growing industrial relevance of QTs,  they are also increasingly discussed in the public media, communicating `breakthroughs' or `quantum advantage' or `quantum supremacy' when claimed by a company. 
In such articles or videos there is a risk of hype, of presenting quantum computing in an exciting but unrealistic way:

\myquote[10~\pureQT, Pos.~26]{It's really important to \textbf{understand what's hype and what's reality.} 
It's not like to get some of the buzzwords and then go in a completely wrong direction. 
[...] quite sometimes people [...] are excited about Quantum, but excited about the wrong thing.%
}%
The term ``hype'' was mentioned 18 times in ten interviews by participants from companies of all types, usually in the context of quantum computing [e.g. 23~\onlyQcomp, Pos.~12; 34~\sme, Pos.~10]. The problem of hype is also discussed as relevant for expectation management of (potential future) employees [28~\pureQT, Pos.~20; 30~\larMulti, Pos.~37].

\subsection{Targeted further training needs in QT}
The interviews revealed different needs for different groups. They may all start with a basic QT awareness as described above. However, they will develop in different directions for further training. 

One group is certainly the \textbf{engineers and technicians} who work around the building of QT. They need specific skills depending on the hardware or software they are working on. These skills are mainly `classical' and are needed in other fields in a similar way [e.g. 12~\pureQT, Pos.~11/quote on p.~\pageref{quote:systemArchitect}; 25~\sme, Pos.~22]. Thus, other fields that require similar skills may be a source for talents:

\myquote[13~\larMulti, Pos.~28]{When you do electronics, very detailed things and you need to put cables in a very confined environment [...] 
because this is like a jewel, a quantum machine. And so looking for this type of profile, the closest field we find, it's like watchmaking or jewellery, where we 
find this very fine technicians.}%
Addressing this will require minor upskilling of `classical' people to `speak the same language' and get an overview, including an idea of what colleagues are doing [e.g. 20~\larMulti, Pos.~21; 18~\pureQT, Pos.~16]. In addition, some may require concrete QT-related qualifications which \textbf{--} due to the lack of technical standardization \textbf{--} are currently very specific to each company, so internal training and learning on the job will usually be necessary [13~\larMulti, Pos.~36-38].

Another group are \textbf{marketing and sales people}, who also require internal training to gain a detailed understanding of their own product. Nevertheless, a general introduction may be beneficial for them, as it would provide a foundation for more detailed training: 

\myquote[02~\onlyQcomp, Pos.~47]{\label{quote:salesAsk}The normal way would be they [i.e. marketing, sales people] ask me, [...] they don't need the in-depth training. Maybe a day or two, just `this is quantum technologies, this can be achieved by it'. But [... they] have to be completely sure what the company does in this area. So external schooling would be very overview like. [... They] will have a talk afterwards with the company guys in the area to say `this is what we are doing, this is what you can sell or promote'.}%

\noindent
A third special qualification may be relevant for people in a department like logistics where e.g. quantum optimization may bring an advantage:\nopagebreak

\myquote[07~\onlyQcomp, Pos.~8, 20, 24]{\label{quote:overseeProduction}%
If you're working in some business unit and you're dealing with certain problems 
[...] if you would then realize that this part is a specific type of calculation that could benefit from a quantum computer [...] and then go to some experts.
[...]\newline
\textbf{Seeing that this is something for a quantum computer is basically the skill} that we would 
like people who oversee production to have. [...] 
there are so many optimization problems where people just are not thinking about quantum computers at all. 
[...]\newline
Maybe two people per factory. [...] 
not everyone working at [company name] has to deal with quantum computers or even has to look at where you could optimize something. But there are people doing that. And those people should know a little bit about it as part of their training. 
}%
People that \textbf{recognize potential use cases} in their own field of work may be relevant in a lot of industries, as there are a lot of fields with potential applications of quantum computing. This may be covered through short training programs, or ideally included into the `classical' qualifications, e.g. study programs, that those people follow [07~\onlyQcomp, Pos.~26].
In addition, one company reports that it has set up a consulting and training department, combining experts from both disciplines, consulting and QT. They educate (potential) customers and help them identify their most relevant use cases. [24~\pureQT, Pos.~9, 12]

With QTs, there is also a need for new bridge roles, people who know the QT and also the business aspects, who read strategic reports and relate them to the own company, and also communicate about QT within the own company:

\myquote[23~\onlyQcomp, Pos.~37]{\label{quote:analyTransl}%
A \textbf{quantum analytics translator}\footnote{The title refers to the McKinsey article on the role of an analytics translator~\cite{henkeAnalyticsTranslatorNew2018}
.} [...] able to understand both worlds [i.e., business and techies] and tell industry, different industries, `this is how we should start and this is why we should start like this way [...] and this is our analytic strategy to make money with quantum computing'.
}%

\noindent
The last special group are \textbf{software designer, data scientists etc.} using quantum computing on a high level:

\myquote[09~\onlyQcomp, Pos.~20]{You have to learn a new library, new type of thinking and so on in the machine learning world anyway. So then to at some point use quantum algorithms and exchange some of the other algorithms for those. [...] we would need courses there, but [...] data scientists already know how to quickly learn a new concept and new languages.}%
Those engaged in high-level quantum computing will require training in the use of the new quantum programming packages and how these differ from `classical' approaches. This may be comparable to the learning of other new packages or languages. Some may require additional courses \textbf{--} or courses within a study program \textbf{--} on the special mathematics and new types of problems that can be addressed using quantum computing [e.g. 21.~\pureQT, Pos.~39; 27~\larMulti, Pos.~19-27]. For the majority of users, the physical realization of the qubits and corresponding quantum concepts will not be essential to include in training in the long term, while currently they are [31~\onlyQcomp, Pos.~26].    

\subsection{Training formats for upskilling in QT (not self-learning)}\label{subsec:trainFormats}

\myquote[32~\pureQT, Pos.~19]{\label{quote:advancedExpertsGap}
Our advanced experts 
don't have the time and capabilities to do in-house training and guide someone as a mentor for weeks or months. Especially if that person comes from very far, [...] 
the gap is too big and 
their agenda is too full [...]. 
I see a difference in what is the kind of the starting point. Are they novice, intermediate or expert? If you're an intermediate, you can follow an expert. But if you're a novice, you need to start from a different starting point.}%
Many of the interviewees, representing a diverse range of companies, report on various forms of internal training and mentoring, as well as onboarding programs that involve sending new hires to a laboratory for a number of days [e.g. 13~\larMulti, Pos.~7; 29~\pureQT, Pos.~40; 32~\pureQT, Pos.~27]. Learning on the job is an important method for providing employees with the specific skills required by the company [e.g. 21~\pureQT, Pos.~24]. For novices to benefit from the expertise of an experienced professional, they must have already reached an intermediate level of proficiency; for those who are less experienced, the gap between their current abilities and those required by the expert is too big [quote above]. 
In addition, advanced people may be send to external courses to become experts [32~\pureQT, Pos.~26].

In particular, start-ups in the process of scaling up have reported the implementation of regular exchange meetings, with one individual presenting their work, a recent publication, or similar, typically on a weekly or monthly  basis [e.g. 28~\pureQT, Pos.~43]. In this manner, employees are able to gain insight into the activities of their colleagues and identify potential sources of support should an issue arise:

\myquote[18~\pureQT, Pos.~21]{We have 
a weekly sort of seminar that's given internally. So everybody gets the opportunity to learn something about somebody else's discipline. 
So my background is not quantum at all. [...] 
I've learned quite a lot in two years from just being inside the business and talking to my colleagues. 
That seems to work quite well.}%
Additionally, hands-on training and workshops are relevant for the purpose of training specific skills and enabling individuals to conduct practical tasks, e.g. working with a specific device or for programming [17~\pureQT, Pos.~32; 11~\onlyQcomp, Pos.~30]. For others, conferences and networking activities are important in order to get updates on research progress or industry needs, as well as for the contacts, e.g. for sales personnel [34~\sme, Pos.~35, 15~\larMulti, Pos. 49-55].

University courses/seminars and online courses are each discussed by about a third of companies, but almost none of the SMEs. Online courses such as MOOCs (Massive Open Online Courses) are preferred by one large company with trainees in different locations [19~\larMulti, Pos.~21]. On the other hand, the advantages of face-to-face seminars, e.g. at a university, are discussed. These allow for more in-depth discussions, practical and laboratory training and supervision, e.g. for algorithm development [27~\larMulti, Pos.~33, 22~\onlyQcomp, Pos.~51].

Another approach is self-learning, Sec.~\ref{subsec:selflearn} discusses different formats for self-learning such as books, podcasts and videos.

\subsection{Follow-up survey results on training needs}
\begin{figure}[ht]
    \centering
    \includegraphics[width=1\textwidth]{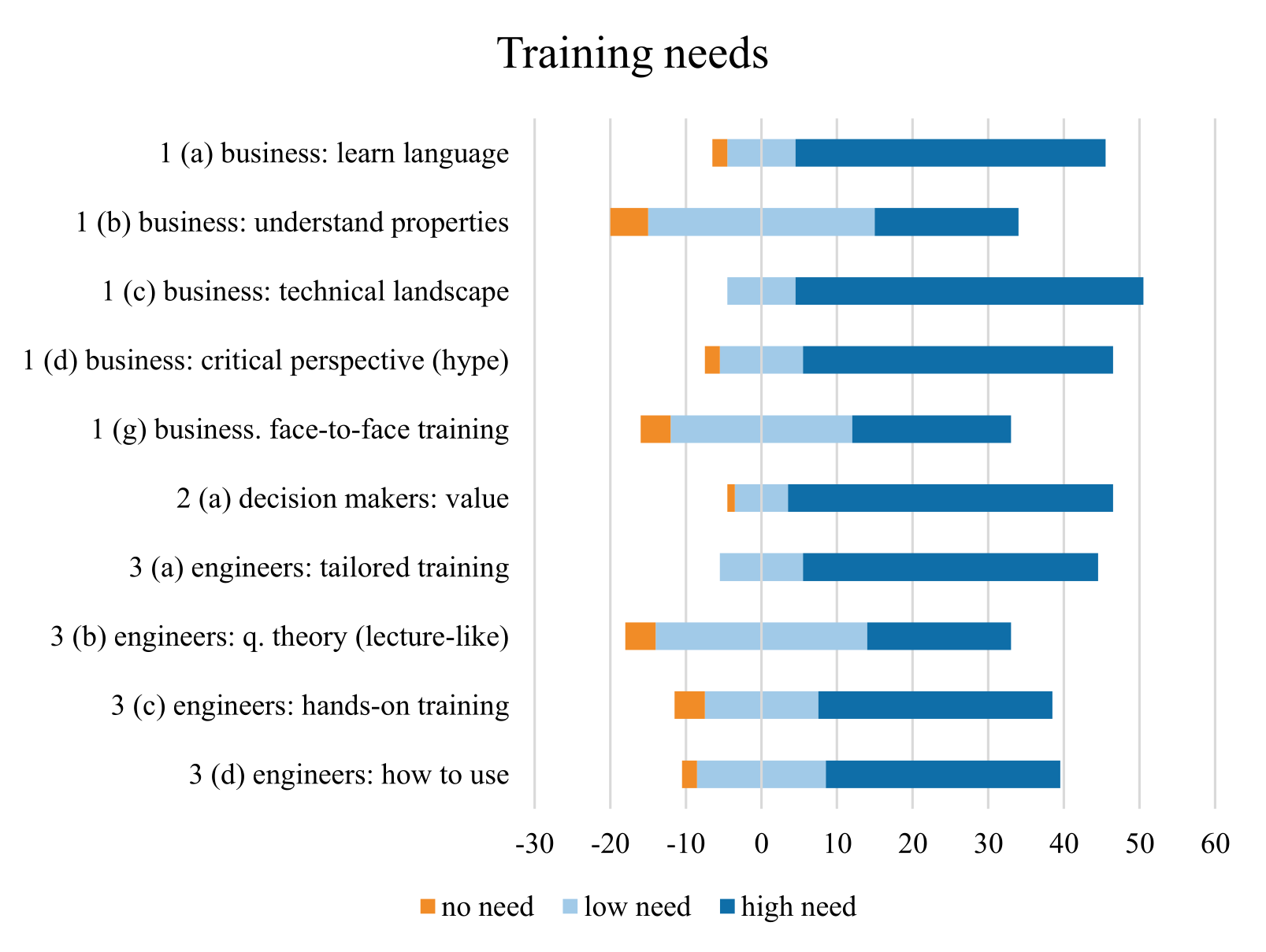}
    \vspace{-0.5cm}
    \caption{Diverging Stacked Bar Chart (DSBC, see Sec.~\ref{subsec:DSBC}) for the training needs from the follow-up survey with 55 participants. 
    Full item descriptions and response statistics are available in Appx.~\ref{subsec:items}.}
    \label{fig:trainingNeeds}
\end{figure}

Figure~\ref{fig:trainingNeeds} shows the response statistics from the follow-up survey on training needs for different groups.

\begin{enumerate}
    \item For \textbf{business} people and colleagues of quantum experts the
highest need is to 
\begin{itemize}
    \item [(c)] \textit{have an idea of the QT landscape, the (expected) applications and use cases, and relevance to their own business}. 
\end{itemize}
In addition, they need to 
\begin{itemize}
\item[(d)] \textit{develop a critical perspective (hype, realistic picture of (im)possibilities, timelines)} 
and to 
\item[(a)] \textit{`learn the quantum language' (without math/physics details), e.g. to enable efficient communication}. 
\end{itemize}
Less needed is that they (b)~\textit{understand the `properties' of quantum, e.g. superposition and entanglement,} or get
(g)~\textit{face-to-face training, opportunities to talk with peers and ask `stupid' questions}. However, even for these aspects, there are about four to five times as many participants who state a high need as those who state no need.

\item For \textbf{decision makers}, politicians, etc., there is a high need to \begin{itemize}
    \item[(a)] \textit{get easily accessible information about potential use and value \textbf{--} without any math or physics}.
\end{itemize}

\item For \textbf{engineers} to go deeper, there is a high need for 
\begin{itemize}
    \item[(a)] \textit{tailored training}, either in
\item[(d)]\textit{how to use a quantum device and integrate it into their work},
or 
\item[(c)] \textit{hands-on training in working with quantum hardware (gain lab experience)}. 
\end{itemize}
Less needed but still relevant is (b)~\textit{lecture-like training or materials on quantum theory, with details on the physics behind QT and advanced mathematics}.
\end{enumerate}

\subsection{New QT master programs and QT experts}
So far we have mainly discussed roles of people with `classical' qualifications who will need some additional QT-specific qualifications, e.g. through upskilling.  
The other group are the new QT experts, people with long-term training and experience. They will also be needed in the QT industry \textbf{--} and this demand cannot and should not be met by quantum physicists alone.

\myquote[22~\onlyQcomp, Pos.~18-19]{\label{quote:differentMindset}%
We are seeing a lot of masters growing everywhere. The problem is, the master is not enough for creating the new professionals skill set. We are today pushing universities for creating degrees on quantum engineering. [...] I hope that the degrees will arrive at the same time that the explosion of the needs will happen in the markets [...] And that is going to create a baseline of people [...]. We will not be able to reskill massively that [old classical] people to the new environment because the paradigm is totally different and we need a different mindset of that professionals.}%
New quantum engineering programs, e.g. master's degrees, will become increasingly important, preparing future employees with quantum concepts and physics fundamentals, as well as strong engineering skills and an idea of how QT can be used. Also here, practical training and student interaction is essential [22~\onlyQcomp, Pos.~51].
Within these programs there are different engineering specializations that are of particular importance for the industrialization of QT, especially systems engineering:

\myquote[30~\larMulti, Pos.~9, see also Pos.~18]{\label{quote:QTsystemEng}
Systems engineering skills [...] 
for large system integrators, 
somebody who is able to understand the different technology bricks to judge, to bring this together 
[...] 
with the very detailed knowledge of quantum physics. 
We call this quantum technology systems engineering 
[... and] this is a very, very big gap because at least to my understanding, there is nearly no university 
offering this kind of skill set today [... covering] the aspect on the system level.}

\myquote[17~\pureQT, Pos.~23]{We are experts in developing and building the hardware, but how they will be set in an infrastructure, [...] how that will interface with the network 
is a very big topic that we have to learn.
}%
Systems engineering and the integration of quantum systems into existing infrastructures is a major challenge within the industrial readiness of QTs \textbf{--} for large integrators as well as start-ups \textbf{--} and something that quantum physicists are usually not qualified for. There is a need for experts with such deep engineering skills as well as a profound quantum physics background.

However, a degree or other (academic/general) course is not enough to be able to work independently in a job. Additional industrial experience is always needed:

\myquote[22~\onlyQcomp, Pos.~65]{
[...] understand that the people 
attending a general purpose course are not 
totally ready for start working in a bank or 
an insurance company or in a logistic company or in utility, and being 100\% productive the first day because they are missing [...] 
the industrial knowledge. 
}

\subsection{Further steps: Qualification Profiles within the European Competence Framework for Quantum Technologies}\label{subsec:qualiProf}
\begin{figure}[ht]
    \centering
    \includegraphics[width=0.75\linewidth]{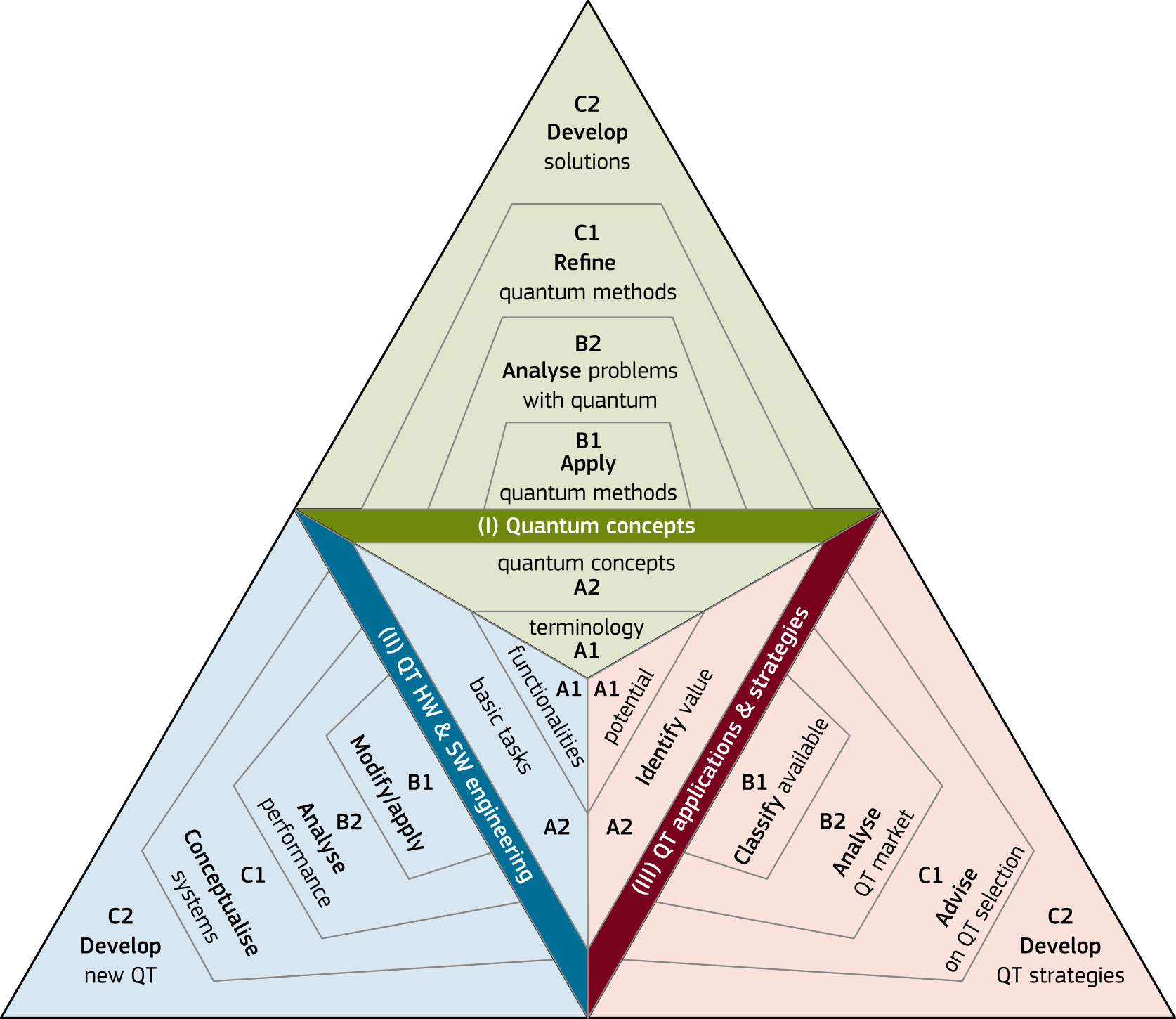}
    \caption{Proficiency triangle from the \CF. In the framework document, each proficiency level (A1 to C2) is described for each of the three proficiency area with an operationalized statement (i.e. starting with verbs such as those in bold in the graphic) and a specification of knowledge and skills. The graphic shows the very short version, just the key point for for each level.}
    \label{fig:profTriangle}
\end{figure}

The findings from the interviews have been  used to update the \CF\ with new Qualification Profiles\footnote{A beta version of Qualification Profiles was already published at \doi{10.5281/zenodo.6834685} (2022). These old profiles are superseded by the new ones as part of the framework update. The new profiles are NOT based on the old profiles and follow a different approach to specifying the qualification, in particular the new profiles are independent of the concrete QT.} 
in spring 2024. 
Key competences identified in the interviews were used to formulate six proficiency levels for the three proficiency areas: 
\begin{itemize}[leftmargin=1cm]
    \item[(I)] Quantum concepts, 
    \item[(II)] QT hardware and software engineering, and 
    \item[(III)] QT applications and strategies. 
\end{itemize}
They form the new proficiency triangle (Fig.~\ref{fig:profTriangle}), which can be used to specify QT-specific qualifications independently of the concrete technology, and thus provide the basis for formulating and visualizing the qualification profiles.

The identification of these qualification profiles started with a sorting based on the material used in the interviews (scheme of competence types, Fig.~\ref{fig:CompTypes}). It turned out that there are more roles than those covered by the material, and that there is some overlap in its initial structure. 
The identification of the roles discussed in the interviews and the associated training needs led initially to eleven roles with corresponding key competences. In an iterative process, including further discussions and expert consultations, the roles were refined to the nine qualification profiles described in the framework version 2.5:
\begin{itemize}
    \item[P1] QT aware person, level A1 in all three proficiency areas
    \item[P2] QT informed decision maker, only A1 in area (III)
    \item[P3] QT literate person (QT literate business role, advocator, enthusiast), A2 in all three areas
    \item[P4] QT practitioner (working with QT, technician, QT user), A1 in areas (I) and~(III), B1 in area (II)
    \item[P5] QT business analyst, A2 in (I) and (II), B2 in (III)
    \item[P6] QT engineering professional (e.g. QT engineer, quantum computer or information scientist), B1 in (I), B2 in (II), A2 in (III)
    \item[P7] QT (HW or SW) specialist (e.g. senior QT engineer, QT architect), B2 in (I), C2 in (II), B1 in (III)
    \item[P8] QT (product) strategist (e.g. advisor, business development expert), B2 in (I) and (II), C2 in (III)
    \item[P9] QT core innovator, C2 in (I), B1 in (II), A2 in (III). 
\end{itemize}

\noindent
To describe the qualification of an individual or the aim of a training, the qualification profile \textbf{--} or more general, the coverage of the proficiency triangle \textbf{--} needs to be combined with a topical selection from the content map, the other part of the \CF. Two examples are provided within the framework document. 

In addition to the qualification profile description, i.e. the desired qualification for a person with example personas, comments are also made on how to achieve this qualification, e.g. what kind of training programs are suitable. They are based on the results of the interview analysis and the follow-up survey. In a way, they summarize and extend some of the results presented in this paper.
Further details of the framework update process and iterative profile refinement will be documented in a separate publication, as they are beyond the scope of this analysis.

\section{Conditions, self-learning formats and more}\label{sec:resultsThree}
This section discusses the findings related to RQ~3~\textit{\ROfour}. They are based on the interview analysis and the follow-up survey. The section is structured into general conditions, details about self-learning materials and formats, and further suggestions collected from the last interview question for an advice from industry to educators.

\subsection{General conditions for training}
The most appropriate \textbf{language for training} depends on the trainees. For engineers or technicians, English is usually fine [e.g. 33~\pureQT, Pos.~30]. As QT is very international and quantum teams are typically international, QT training for engineers may need to be in English [16~\pureQT, Pos.~30]. 
In the interviews, about 70\% (18 of the 26 interviews discussing this question) said they prefer English. For several companies, English is the company language, so employees need the vocabulary in English [e.g. 32~\pureQT, Pos.~29]. However, even if English is preferred, in some countries a translation to the local language may be required by law [20~\larMulti, Pos.~35]. 

With advances in machine translation, an automated translation process may work in the background for videos or other materials to increase accessibility for those who prefer the local or native language [e.g. 17~\pureQT, Pos.~44; 31~\onlyQcomp, Pos.~34]. This may be particularly important for the older generation or non-technical people, e.g. operators or politicians [e.g. 31~\onlyQcomp, Pos.~36] \textbf{--} or for the first exposure to quantum concepts, where they are likely to be complicated and English may be a barrier, making learners think more about the words than the concepts [16~\pureQT, Pos.~30]. 
Another reason for using the local language is to build trust:

\myquote[17~\pureQT, Pos.~34]{We see that trust and security also relies on a common language, which is very much known to them. So we want to convey, of course, more than just selling a piece of hardware and say `do whatever you wish', but also convey a sense of security. And that's why this common language also can help a lot.}

\begin{figure}[ht]
    \centering
    \includegraphics[width=1\textwidth]{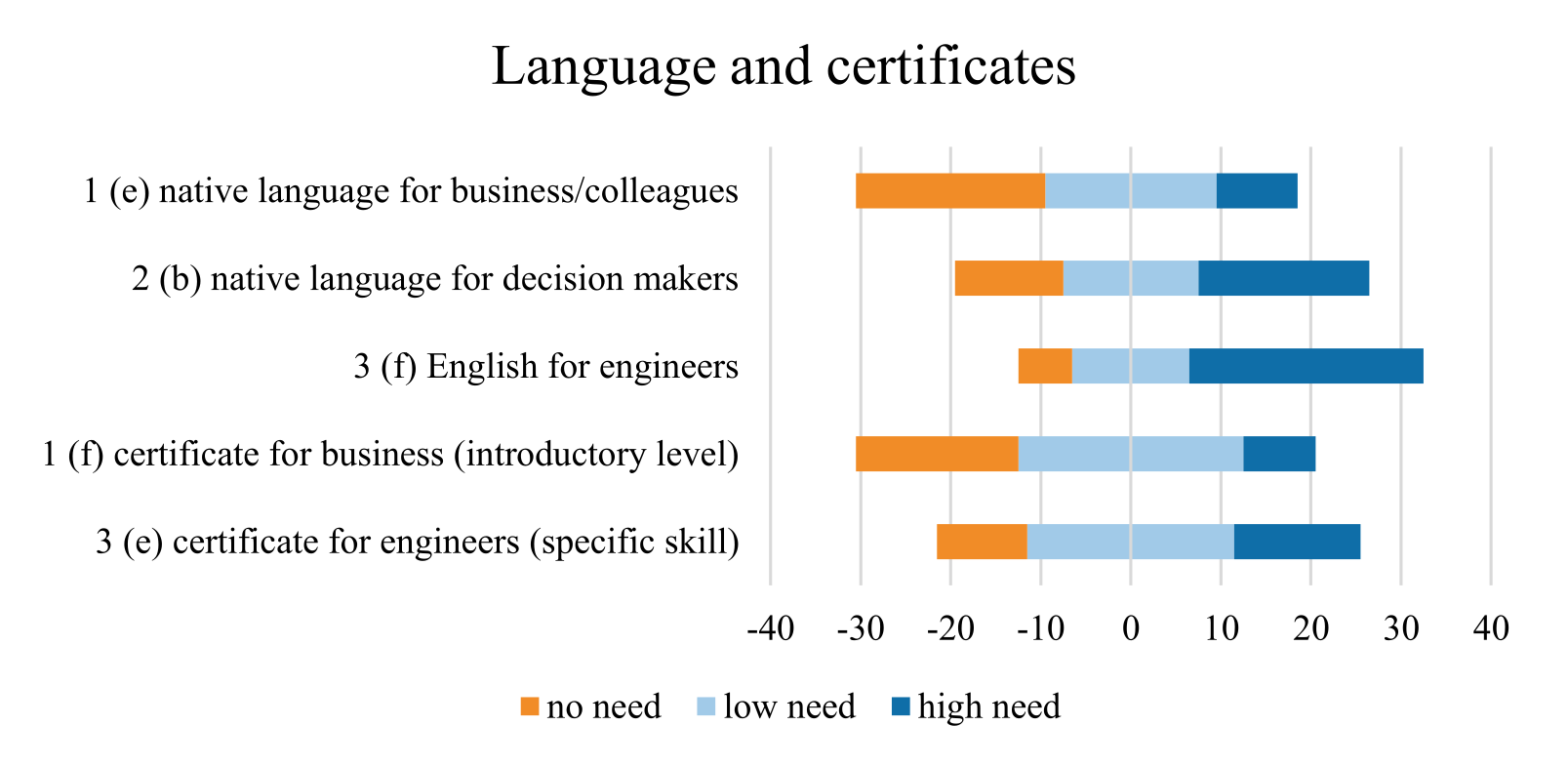}
    \vspace{-0.5cm}
    \caption{DSBC showing the needs for having training in native language or English as well as for certificates after a training, visualization of the responses from the follow-up survey with 55 participants.
    Full item descriptions and response statistics are available in Appx.~\ref{subsec:items}.}
    \label{fig:langCert}
\end{figure}

\noindent
The follow-up survey included three items on language needs. Although the opinions were very heterogeneous, there were some tendencies, see Fig.~\ref{fig:langCert}.
There is rather no or low need 1~(e)~\textit{for business people and colleagues of quantum experts to have training in the native language, as this makes it easier to understand the principles, rather than having everything in English}. In comparison, there is a higher need 2~(b)~\textit{for decision makers, politicians, etc. to have it [i.e. easily accessible information on the potential value creation with QT] in their native language to facilitate understanding and increase trust.}
The highest rated need is 3~(f)~\textit{for engineers to go deeper to have everything in English to communicate internationally.} However, the number of respondents who did not rate these items is noticeable higher than for other items (see Tab.~\ref{tab:ratings}).

\label{subsec:certificate}

\noindent
The second condition discussed in the interviews and also assessed in the survey is the need for \textbf{certificates} for the training. Fig.~\ref{fig:langCert} also shows the responses to the two items on certificates: While they are less needed for 1~(f)~\textit{business people and colleagues of quantum experts at an introductory level}, the need increases for 3~(e)~\textit{engineers after attending a course in which a specific skill is trained}.

Also the interview analysis indicates that certificates become more important when a specific skill is being trained, whereas for some basic awareness a certificate of attendance should usually be sufficient [e.g. 08~\pureQT, Pos.~53-57]. Getting some kind of certificate can be important for the CV for a later job application \textbf{--} if it is really reputable; and also within the current job to show to HR people/management and to justify the time and money spent on the training to the company [e.g. 07~\onlyQcomp, Pos.~51; 13~\larMulti, Pos.~59]. Some interviewees indicated that this is less important in start-ups as there is a different kind of social dynamic in the team, where it will be visible anyway if a training shows progress in the work of the participants [e.g. 16~\pureQT, Pos.~32].

Another argument for certificates mentioned in the interviews is that it can be relevant for motivation [e.g. 17~\pureQT, Pos.~33]. However, it is important that it comes from a trustworthy institute or authority [e.g. 13~\larMulti, Pos.~59], and is somehow standardized\footnote{ 
Providing a reference for QT education with a certification scheme based on the \CF\ is one aim of the QUCATS project and a step towards standardization of QT education. This relevance is also discussed in [17~\pureQT, Pos.~33].}:

\myquote[33~\pureQT, Pos.~32]{I think it is always better to have a standard. So 
the question is how to standardize these trainings? 
[...] It comes back to the ECTS systems from 
education. I mean, they are standards. I think there should be an audit from the authority that comes and says 
`indeed, it is a good training'. 
}

\noindent
Nine interviews provided estimates of the reasonable \textbf{cost of training}. 
The answers range from a few hundred to about a thousand euros per day for face-to-face training, depending on how much practical work is included [e.g.~01~\sme, Pos.~36]. Respondents note other costs of sending people on training (travel, hotel, working time), so the fee for the training may not be the largest contributor to the total cost of sending people on training [08~\pureQT, Pos.~61]. 

In principle, there could be huge differences depending on the country and the size of the company, as the cost of a working day per person also varies between countries [12~\pureQT, Pos.~47] 
and target groups, e.g. students and job seekers, i.e. people without a company paying for them [09~\onlyQcomp, Pos~35]. However, while a lower price may be a strategy to reach more people, a free course may also be sub-optimal:  

\myquote[12~\pureQT, Pos.~45]{%
Maybe a 
small participation fee is good to select a minimum the participants. Otherwise, 
if you do this in a city, you get a lot of people from that city just because it's for free.%
}%
Another suggestion would be to start with a free overview course online to attract a larger group of people, followed by more advanced paid courses [12~\pureQT, Pos.~48]. 

\subsection{Self-learning}\label{subsec:selflearn}

\myquote[34~\sme, Pos.~21]{%
What I do see, especially with a slightly older workforce, is that quantum \textbf{--} because they cannot grab it, it's a bit scary. 
[...] the self-learning training or at least the  planting of seeds kind of thing, that it is not that scary. It's just a different way of thinking.}%
Even for self-learning materials, one format does not fit every need. A combination of formats is required and, depending on the qualification objective, self-learning may need to be combined with personal training e.g. 16~\pureQT, Pos.~18-20]. Furthermore, the most appropriate format may depend on the generation of learners and the situation in which they find themselves in [e.g. 17~\pureQT, Pos.~44]. 

This section discusses different formats for self-learning materials, ranging from podcasts to books and videos, as well as more general needs such as the relevance of interactive formats or support in finding materials. One aspect considered important for self-learning materials is easy and offline accessibility, so that they can be used quickly between meetings or, for example, on a plane without internet access [e.g. 29~\pureQT, Pos.~54; 07~\onlyQcomp, Pos.~32].

~\newline\noindent
\textbf{Podcasts} were mentioned in eight interviews and are consumed, e.g., during the commute to work [32~\pureQT, Pos.~21]. 
Although podcasts are deemed appropriate to provide an overview of 
trends and expectations, they were not considered the optimal format for complex concepts that require visualization, exercises, or hands-on engagement [quote below; 17~\pureQT, Pos.~32]. Furthermore, the absence of interactivity was thought to result in listeners retaining less information [07~\onlyQcomp, Pos.~32].

\myquote[03~\sme, Pos.~43]{%
It depends on the concept that's being discussed. It can be a bit more difficult to follow. [...]
If you're talking about general market trends and things like `this is what we expect to happen', then a podcast is extremely useful. If you're trying to convey 
fairly complex physical concepts, that's difficult to do in a podcast.
}%

\noindent
\textbf{Video courses and clips} were discussed in several interviews.  
YouTube, for example, offers a vast array of videos, with some individuals utilizing it to watch university lectures [03~\sme, Pos.~38] or learn about a quantum programming language with a recorded summer school [30~\larMulti, Pos.~31]. In general, YouTube videos are reported to be used for awareness or to refresh one's knowledge [13~\larMulti, Pos.~49-51; 19~\larMulti, Pos.~35]. 

Some companies report having an internal playlist or learning platform, whereas one interviewee notes a problem with publicly available videos: 

\myquote[20~\larMulti, Pos.~31]{Well, we cannot use public videos from the external world. That's one rule. We need to buy [videos] to integrate into our catalog. 
We also develop 
[...] 
and this [videos] we can 
put in our playlist available to the employees.
}

\noindent
\textbf{Interaction} is recognized as important for reinforcing learning, especially when it goes beyond short single-choice questions:

\myquote[21~\pureQT, Pos.~35]{
Some sort of feedback reinforcement on the learning is key because [...] 
a podcast or a video without that, it's very easily gone within a very short time. So, 
working through exercises 
is probably actually the best in terms of 
having to deliver something that really reinforces the learning. 
[...] 
The value of 
short questions is somewhat limited. But 
if people are actually getting in there and have to work through the concepts themselves, then that really drives home the learning.}%
In contrast, mandatory exercises can also be a barrier for those who want a quick overview and don't want to spend time on exercises, so exercises should be optional, perhaps only required if someone wants to get a certificate [10~\pureQT, Pos.~28].

\begin{figure}[ht]
    \centering
    \includegraphics[width=1\textwidth]{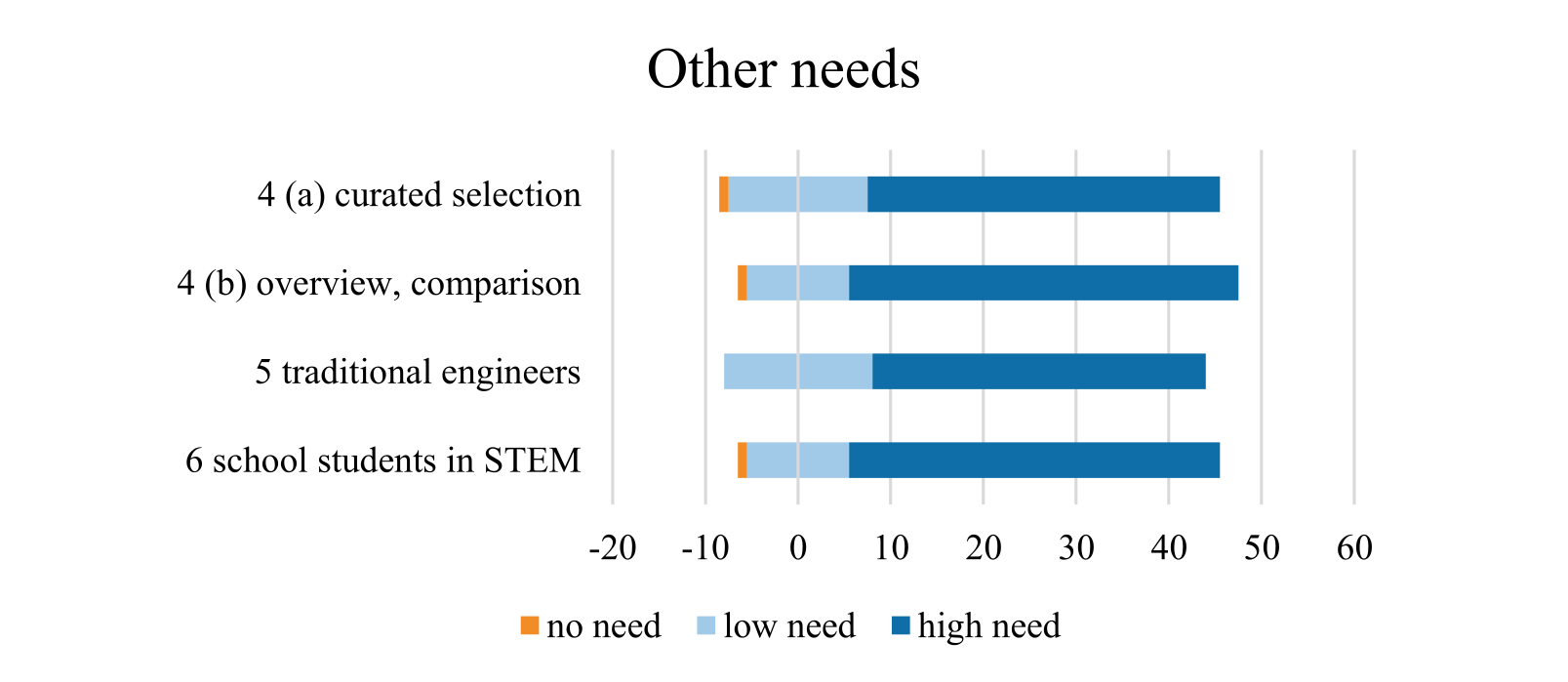}
    \vspace{-0.5cm}
    \caption{Other needs assessed in the follow-up survey. Full item descriptions and response statistics are available in Appx.~\ref{subsec:items}.}
    \label{fig:otherNeeds}
    \vspace{-0.5cm}
\end{figure}

~\newline\noindent
The interviewees indicated that \textbf{written text} is an important reference tool \textbf{--} but may be less engaging than videos, podcasts, or similar formats for the majority, particularly for long texts like books [03~\sme, Pos.~43; 07~\onlyQcomp, Pos~30; 32~\pureQT, Pos.~20]. Some individuals prefer to get started with a book, before going into practical tasks, hands-on training or similar [11~\onlyQcomp, Pos.~30]. 
However, some formats of written text were mentioned as interesting:
\begin{enumerate}
    \item  An informational web page similar to Wikipedia, a one-page document with a list of advantages and disadvantages and a brief summary [11~\onlyQcomp, Pos.~20]. However, there is a risk that it will become outdated quickly, necessitating regular updates, thus it may be beneficial to consider including this as a handout in training sessions or on the Quantum Flagship website~\cite{qucatsQuantumFlagshipFuture2024} [12~\pureQT, Pos.~57].
    \item PowerPoint slides with summaries for management:\nopagebreak 

    \myquote[11~\onlyQcomp, Pos.~20]{
    People at the management level 
    are very busy, [...] 
    the shorter the better. So 
    the material what I would suggest 
    this one page overview, pros, cons, use
    , maybe short video clips. 
    These people do like PowerPoint presentations, 
    short, 
    with the basic effects. No details, trust summaries. 
    [...] They are not interested how many qubits there are on the systems they have. They are interested how many money I can gain with this algorithm, for example.
    }
    \item A blog that separates hype from reality, provides summaries and breakthroughs, and is written by a good communicator so that it is accessible for business people [10~\pureQT, Pos.~31]. 
    \item Concise script at the end of a training:
    
    \myquote[16~\pureQT, Pos.~53]{
    have in the end documents, really 
    documents to read, not like books and not say, `hey, you can read this book' because nobody will do it. 
    [... Work together in the seminar on] a really good script with some equations, easy equation and really good legends. Really good explanation.}
    \item Scientific documents, paper and in particular 
    reports such as the Quan\-tum Flagship´s Stra\-te\-gic Research and Industry Agenda (SRIA, available at \cite{qucatsQuantumFlagshipFuture2024}):

    \myquote[12~\pureQT, Pos.~23]{
    The SRIA, the strategic report 
    gives already a good overview. 
    Talking about 
    people 
    considered to be high profile, 
    it´s a good exercise to let them study documents coming from the flagship, 
    from QuIC [...]
    }
\end{enumerate}%
The high need for short overview pages, more concretely 4~(b)~\textit{a brief overview of quantum technology approaches and available solutions with comparison (pros \& cons)} was also confirmed in the follow-up survey, see Fig.~\ref{fig:otherNeeds}.

\subsection{Support in finding materials}
Another follow-up survey result is the high need of 4~(a)~\textit{a curated selection of good online material on QTs and quantum theory (YouTube videos, podcasts, blogs,~$\dots$)}, see Fig.~\ref{fig:otherNeeds}. The item was formulated based on this need expressed in the interviews:

\myquote[32~\pureQT, Pos.~49]{I think doing that on your own on Google is like a needle in a haystack. There's too much available. And 
you're unable to judge `what is too advanced for me versus what is not advanced enough versus what is priority or not'. So 
having some form of a portal that 
helps you, this is what matters.
}

\myquote[06~\larMulti, Pos.~24]{There's a lot of material available. 
I think it's good if this can be structured somehow. So you put it into different symbolic boxes or so and you check if it fulfils quality criteria. 
And then maybe you would have a web page or so with different categories and there are different links that lead to some YouTube videos or so, that's fine. I don't think we have to develop this stuff from scratch.}%
Respondents see that there is a lot of good material available, but that it is a challenge to find the right material for an individual. They see a need for a platform where a user can specify their previous qualification, where they are starting from, and their goals, where they want to go [e.g. 18~\pureQT, Pos. 25, 45]. And then they get a suggestion of good quality materials, that quality control is really important [e.g. ibid, Pos.~45].

\subsection{Additions and non-quantum needs}
There are some other results from the interviews and the follow-up survey. From the latter, we see a high need to  (5)~\textit{get more traditional engineers interested in working in the quantum domain} and an even higher need to
(6)~\textit{get more school students interested in STEM and quantum}, see Fig.~\ref{fig:otherNeeds}.

And there are also further needs outside the quantum-related field of education, as collected towards the end of the interviews:

\myquote[08~\pureQT, Pos.~69]{The level of education and training that we actually make use of in a company of a person that has four years of education in physics or quantum physics is very limited. 
[...] the education really is about problem solving and how to [...] 
plan your activities, how to be a professional in the sense that you're able to execute tasks that are complicated. 
It's more about learning how to learn, 
[...] 
how to be efficient, 
how to get things done. And also assessing the quality yourself. }

\myquote[17~\pureQT, Pos.~62]{The main problem globally is the bad education from mathematics. I mean it's not that the people are not talented in mathematics, but the education for mathematics is very low level in the whole world. Why? I don't know.
}%
Beside \textbf{--} and sometimes more than \textbf{--} quantum skills, more general skills like  problem solving, teamwork, knowing the specific resources or communication skills are needed [e.g. quote above, 05~\onlyQcomp, Pos.~42, 57]. Some people may need special training on quantum-related mathematics, and bad education in mathematics was mentioned as a problem [quote above; 21~\pureQT, Pos.~39].

Another specific training need is business in start-ups, discussed by 6 of 13 start-ups: they have the special challenge that academics and technical people have to cover business/management roles:

\myquote[18~\pureQT, Pos~48]{to succeed in a business, then you need to understand business as well as understand quantum.}%
One interviewee reported on an intensive mentoring program that helped develop their management skills [26~\pureQT, Pos.~26]. 
Some interviewees say that general business training is not appropriate for the specific situation of a high-tech start-up [e.g. 26~\pureQT, Pos.~24], others report a successful combination of business experienced people with technical people in the start-up [29~\pureQT, Pos.~42; 32~\pureQT, Pos.~44]. However, it was suggested that QT training providers should focus on quantum as there is enough general business training available [32~\pureQT, Pos.~47].

\subsection{Industry advice}
The interviews were concluded with the question: \textit{In order to develop materials or trainings that really address the industry, what advice would you give to us educators?} The main findings from this last part of the interviews are discussed below. 

Interviewees from nine companies stress the importance of talking to industry and getting to know the audience of industry training courses, such as the engineers and team leaders:

\myquote[34~\sme, Pos.~51]{Learn about your audience. 
That's the sales guy in me. If you need to sell a new topic to somebody else, you need to step out of your own world and step into the other world to see what their way of working is. And connect your information to that and not vice versa because you need to make it your customer as easy as possible to get this new information to it. And don't assume that people know what quantum is. You can assume that most people find it a bit scary because they don't understand it.}%
They see the importance of linking content to the industry world and also getting industry feedback [e.g. quote above; 28~\pureQT, Pos. 48-49]. Ideally, industry partners should be co-developers, also to gain credibility and reach out to the community [10~\pureQT, Pos.~59]. And industry partners can benefit from a joint effort to train students, job seekers and the like, as this is an opportunity to identify talent and recruit good course participants [23~\onlyQcomp, Pos.~19, 21, 27]. 

However, the association of a course with a particular company can have a bias, a competitive effect, and providing or supporting a training can be relevant for marketing [12~\pureQT, Pos.~53]. It is therefore important, especially as an academic course provider that suggests a balanced perspective, to include several companies and products and to strive for a neutral perspective [ibid, Pos.~55].

In terms of targeted company types, QT end-users and large, old companies seem to have the higher demand for training, so they should be involved in the training preparation to really meet their needs [14~\sme, Pos.~47-49].

~\newline\noindent
There are some more suggestions collected in the interviews. 
They appear only one or very few times in the interviews, but were identified as interesting for the development of training or educational materials, for example:

\myquote[11~\onlyQcomp, Pos.~46]{It should be concise, 
focused, 
interesting and modern. And it should be clear what are the expectations? What is the outcome? What is it worth to do it? So to address this topic and say we offer a program for upskilling quantum computing in finance and people ask at the end, 
what can we do now? 
It's nice, we know about something, but how can we use it in work? [...] 
If it's possible to make this connection, then it will become interesting for the industry.}%
The first suggestion is to clearly communicate what is expected, what makes the training or course worth doing, and to link this to the audience's world (application-oriented) [quote above; 23~\onlyQcomp, Pos.~29].
Other suggestions are:
\begin{itemize}
    \item Modular approach: some basics plus a variety of advanced courses tailored to specific applications or companies. This can start with an overview, an umbrella story, and then zoom in on specific topics and skills [32~\pureQT, Pos.~46].
    \item Training is more for the basics, experts can read papers etc. [07~\onlyQcomp, Pos.~63] \textbf{--} but for them it may be interesting to get market updates, recent developments and the like in a condensed format, thus keep the experts updated [30~\larMulti, Pos.~14].
    \item Training should allow for engagement and discussion [04~\onlyQcomp, Pos.~72], so face-to-face training may be useful even if no practical skills are being trained.
    \item Materials should be concise (short attention span), available offline (use in downtime) and easy to use, especially for people who `have to' (not `want to') learn about QT [29~\pureQT, Pos.~54; 16~\pureQT, Pos.~52].
    \item Use the language of the target audience, to avoid fear of wording and to offer points of connection [e.g. ibid.].
    \item Innovative, interesting and modern approaches, e.g. competitions, make it fun and an event to get people involved, make them feel that quantum is great and that they want to be trained [e.g. 20~\larMulti, Pos.~55].
    \item Give an idea of what kind (and size) of problems QT can or might solve. Focus on early (potential) use cases and where the greatest or earliest value is expected [e.g. 23~\onlyQcomp, Pos.~21, 27; 31~\onlyQcomp, Pos.~44].
\end{itemize}

\noindent
In addition, one interviewee suggested a platform that brings together developers, e.g. quantum software developers, and industrial end users who don't want to hire their own experts, e.g. in small companies. On this platform, end-users can post a problem, perhaps set a price, and get experts to work on it and find the best solution (of what these experts come up with) in a competitive way. \footnote{The interviewee refers to the platform kaggle \url{https://kaggle.com/}, which follows a similar approach in data science.} 
This is expected to push quantum computing forward with real-world use cases. [23~\onlyQcomp, Pos.~40, 44]

\section{Summary}\label{sec:sum}
In this section, short summaries of the results discussed in Sec.~\ref{sec:resultsChallenges} to \ref{sec:resultsThree} are provided in order of appearance. In general, a large number of points were raised by the interviewees in all three research questions, allowing for a general assessment of the issues. Each research question had a variety of aspects that were raised and elaborated by the participants.\\

\noindent
For RQ~1: \textit{\ROone}
\begin{description}
    \item \textbf{Talent shortage} exists, but maybe not special compared to other technologies, and QT may also attract talents.
    \item \textbf{Lab experience} is important, and is lacking among new graduates due to COVID.
    \item \textbf{PhD} physicists less needed, instead need more engineers \textbf{--} but post-docs are lacking in academia.
    \item \textbf{Internal QT training}, e.g. for engineers, is needed to cover company-specific needs (lack of technological standardization).
    \item \textbf{Upskilling programs}  should focus on the basics/overview, e.g. for engineers, to prepare them to learn on the job.
    \item \textbf{Need quantum easy} accessible for non-technical people.
    \item \textbf{QT aware or literate workforce}, e.g. engineers, are lacking and perhaps scarcer than \textbf{QT experts}, e.g. quantum physicists. 
    \item \textbf{Autonomous further qualification} of some classical experts to become hybrid experts.
    \item \textbf{Quantum programming} starts slowly due to unclear perspective, prepare hybrid experts, mainly in very large companies.
    \item \textbf{Decision makers} in very large companies usually ask their QT experts, but need some awareness.   
    \item \textbf{Different professions} are discussed in different frequencies, most of them in the engineering domain and especially software engineering, but also physicists, other STEM and business roles.
\end{description}
~

\noindent
For RQ~2: \textit{\ROthree}
\begin{description}
    \item  \textbf{QT awareness} on impact, ethics and hype is relevant, also for the broad society.
    \item \textbf{Learning on the job} is done a lot, but need intermediate not novice.
    \item \textbf{External training} to bring novice to intermediate that can learn on the job.
    \item \textbf{In-person} training is important to train hands-on lab skills, and also for discussion.
    \item \textbf{Engineers and technicians} need classical skills and some reskilling, specific for their tasks.
    \item \textbf{Marketing and sales people} may benefit from external courses on foundations, but need product-specific education.
    \item \textbf{People that recognise potential use cases}, in particular of quantum computing, will be beneficial in lots of QT application areas.
    \item \textbf{Software designer} and similar may learn high-level quantum programming, thus new languages and packages, like any other new language or package. 
    \item \textbf{Mixture of formats} required to address different people and preferences, but also for different contents or skills, e.g. external training \textbf{--} online and/or hands-on workshops \textbf{--} plus internal exchange plus learning on the job. 
    \item \textbf{Business people} need an idea of the QT landscape with a critical perspective and to learn the `quantum language'.
    \item \textbf{Decision makers} need to get the value.
    \item \textbf{Engineers} need tailored training with practical tasks.
    \item \textbf{New masters} should integrate systems engineering aspects.
    \item \textbf{Results are incorporated} into  version 2.5 of the \CF, adding a proficiency triangle and nine qualification profiles to the content map from version 2.0.
\end{description}
~

\noindent
And for RQ~3: \textit{\ROfour}
\begin{description}
    \item \textbf{English} is fine for engineers, they need the vocabulary in English.
    \item \textbf{Local language} might be good for an easy entry, to reduce cognitive load, e.g. for non-technical people and if trust plays a role.
    \item \textbf{Certificate} important if a specific skill is trained, otherwise proof of attendance sufficient.
    \item \textbf{Free access} for the introductory materials, subsequent courses may be fee-based.
    \item \textbf{Materials} for self-learning instead of in-person courses for busy people like management, offline available to use in down-time.
    \item \textbf{Podcast} nice for way to work, for e.g. market trends, not for quantum concepts.
    \item \textbf{Videos} rather for the basics, already a lot available.
    \item \textbf{Interactive elements} might be barrier for some people, but some reinforcement important to really learn something.
    \item \textbf{Written text} relevant to have something to refer to, most people do not take the time to read.
    \item \textbf{Support in finding materials} is needed, with quality control, and option to specify aims and prerequisites.
    \item \textbf{Need more students} in STEM and more engineers in QT industry.
    \item \textbf{Industrial co-developers} for training courses/materials to really meet their needs.
    \item \textbf{Clear objectives}, modular approach, concise materials, easy and fun, e.g. with competitions.
\end{description}

\section{Discussion}\label{sec:disc}
For this study, we evaluated 34 interviews with industry representatives from Europe, conducted in summer 2023, and 55 responses to a follow-up survey. We see that the QT industry faces a number of challenges in terms of workforce development. One challenge is the lack of talent. Not only are more experts needed, but new skills are required to turn a laboratory set-up into a product \textbf{--} skills that cannot (only) be covered by quantum physicists. In addition, the currently growing start-ups need new people in `classical' roles with some QT awareness and interest to work in a QT company. 

On the one hand, getting people interested in QT can be a huge challenge, as people tend to be afraid of quantum, finding it `spooky' and incomprehensible. These people need to be reached, and made excited and enthusiastic about QT, so that they want to start learning about QT and work in a QT company. On the other hand, talent may also be attracted by the new opportunities and challenges and therefore want to work on QT. A variety of engineering roles, most often in software, are relevant to the QT industry and are discussed in more detail than physicists, chemists and other STEM roles. The greatest shortage seems to be of engineers with qualifications and experience in QT, or of engineers who are interested in getting into QT.

Fox et. al.~\cite{foxPreparingQuantumRevolution2020} report findings on the types of jobs needed, based on 22 interviews with US companies in autumn 2019. Also in this study, engineers are the most discussed group. However, they are not as dominant as in our analysis, and electrical engineers dominate software engineers, whereas in our interviews software engineers were mentioned more frequently. This may indicate a shift from hardware and electronics to a greater emphasis on software as the industry becomes more mature. We did not explicitly ask to name all roles, but analyzed which roles were named during the interview, assuming that these were the most relevant. This may limit the validity of this comparison. Also Aiello et al.~\cite{aielloAchievingQuantumSmart2021} report of the growing need for engineers with ``traditional'' engineering expertise and some additional quantum education. They discuss how this add-on can be provided through elective courses in degree programs or through dedicated new programs.

We observed that several business-oriented bridge roles are becoming more relevant as the QT industry matures. Already Hughes et al.~\cite{hughesAssessingNeedsQuantum2022} identified business as a third relevant block alongside hardware and software to classify the skills required for specific jobs within the quantum workforce. In that study, they also reported that a PhD is relevant for only a few roles, and that some form of short-term quantum training should be sufficient for several roles.

Our interviews revealed a need for different levels and formats of short-term training. The most basic is a few hours of seminars or short materials such as podcasts, video clips or one-page summaries to raise QT awareness with a focus on impact and hype. 
The follow-up survey revealed a very strong need for business people to grasp the technical landscape and for decision makers to recognize the value of QT. 

Currently, a lot of learning is done `on the job', as already described by Fox et al.\cite{foxPreparingQuantumRevolution2020}. We have heard from several start-ups about their approach of regular internal exchange meeting to educate colleagues. However, some external training on the foundations, to bring a novice up to the level of a (QT literate) intermediate, should be useful, as for a novice the gap would be too big to learn from an expert. 
Regarding training of ``quantum-literate technicians'', there is a study by Hasanovic et al.~\cite{hasanovicQuantumTechnicianSkills2022}. It includes the assessment of 24 participants (75\% from industry) of skills \textbf{--} sometimes rather knowledge \textbf{--} for technicians, and a suggestion for curriculum development. 

Summarizing, we see a need for short-term training of (a) the broad overview to be able to learn more on the job, and (b) dedicated hands-on training, focusing on practical tasks relevant to what employees would have to do with or around quantum technologies. For these two formats, interaction and learning reinforcement are considered essential.

As with dedicated short-term training, such a qualification may also be covered through elective courses within degree programs. Nevertheless, there is a need for new QT (masters) programs with a strong engineering component: especially systems engineering skills are currently lacking in the QT workforce.

While training and materials in English are fine for engineers, others may benefit from having them in their native language, to focus on the content rather than the wording.
Certificates from upskilling seem to be more important for employees to show something to the company, rather than for jobseekers, as they tend not to be taken into account in hiring processes.
There is already a lot of self-learning material available, what is needed is a platform helping to find good (quality controlled) material, with the ability to specify previous qualifications and learning objectives. 

\section{Conclusion and further work}
This paper reports on challenges and strategies for workforce development, needs for training programs and formats, and other suggestions from industry on how educators could support the development of a quantum workforce. 
The results have been incorporated into the \CF\ update to version 2.5, an output of the QUCATS~\cite{qucatsQuantumFlagshipFuture2024} project. With this update, the framework is a tool for planning, mapping and comparing qualifications, training objectives and job requirements in two dimensions: the QT content map and the new proficiency triangle.  
New qualification profiles summarize the typical roles relevant to industry, together with suggestions for training or learning pathways. By the end of the QUCATS project (in 2025), a certification scheme based on the framework will be developed to provide a reference for certificates and make them comparable at European level.

Another output of QUCATS is the `Quantum Playlist', a collection of reviewed videos structured according to the framework. It is hosted at the EQRC~\cite{quantumflagshipEuropeanQuantumReadiness} website, as mentioned already in the introduction.
This is a first step in supporting the search for high-quality materials, as requested in the interviews, but without the ability to get personalized suggestions. 

Finally, at the European level, the QTIndu~\cite{qurecaQTInduQuantumTechnologies2024} project provides modularized online courses (videos, texts, interactive elements) and also some face-to-face training. Examples are online curses on the QT ecosystem or basic quantum effects, a hands-on training on quantum sensors or a workshop for business professionals, HR and the like on skills and competences in hiring and training, using the \CF\ as a tool. 

\section*{Abbreviations}
\begin{itemize}[leftmargin=2cm]
    \item[CEO] Chief Executive Officer 
    \item[CSA] Coordination and Support Action
    \item[CTO] Chief Technology Officer
    \item[DSBC] Diverging Stacked Bar Chart
    \item[DigiQ] Digitally enhanced quantum technology master (project title) \cite{shersonDigiQDigitallyEnhanced2024}
    \item[EQRC] European Quantum Readiness Center \cite{quantumflagshipEuropeanQuantumReadiness}
    \item[EQTC] European Quantum Technologies Conference
    \item[FPGA] Field-Programmable Gate Array
    \item[HR] Human Resources 
    \item[MOOC] Massive Open Online Course
    \item[PR] Public Relations
    \item[QT/QTs] Quantum Technology/-ies
    \item[QTIndu] Quantum Technologies courses for Industry (project title) \cite{qurecaQTInduQuantumTechnologies2024}
    \item[QUCATS] Quantum Flagship Coordination AcTion and Support (project title)\cite{qucatsQuantumFlagshipFuture2024}
    \item[QuIC] European Quantum Industry Consortium \cite{quicEuropeanQuantumIndustry2024}
    \item[SME] Small or Medium-sized Enterprise
    \item[STEM] Science, Technology, Engineering, and Mathematics
\end{itemize}

\backmatter

\section*{Declarations}
\bmhead{Data availability}
The interviews or interview transcripts are not published due to data protection reasons, only the referenced phrases (quotes and other references to the interviews from the results sections) are provided. These and the data from the follow-up survey are available at Zenodo~\cite{greinertSupplementaryMaterialQuantum2024}, together with the full interview guide.

\bmhead{Acknowledgements}
We thank Mattia Giardini (QuIC), who helped arrange many of the interviews, and all the study participants, especially those who took the time to participate in the interviews.

\bmhead{Author contributions}
FG prepared, conducted and analyzed the interviews and follow-up survey and wrote the original manuscript, aided by MU. IND contributed to the reliability evaluation. All authors contributed to the study design and analysis, i.e. they discussed the methods and results, revised the manuscript, and read and approved the final manuscript.

\bmhead{Funding} 
This work  has received funding from the European Union’s Horizon Europe research and innovation programme under grant agreement No 101070193 (FG) 
and from the European Union’s Digital Europe programme under grant agreement no. 101100757 (IND, DHR).
\hfill \raisebox{\dimexpr-\totalheight+\ht\strutbox\relax}{
	\includegraphics[width=1.5cm]{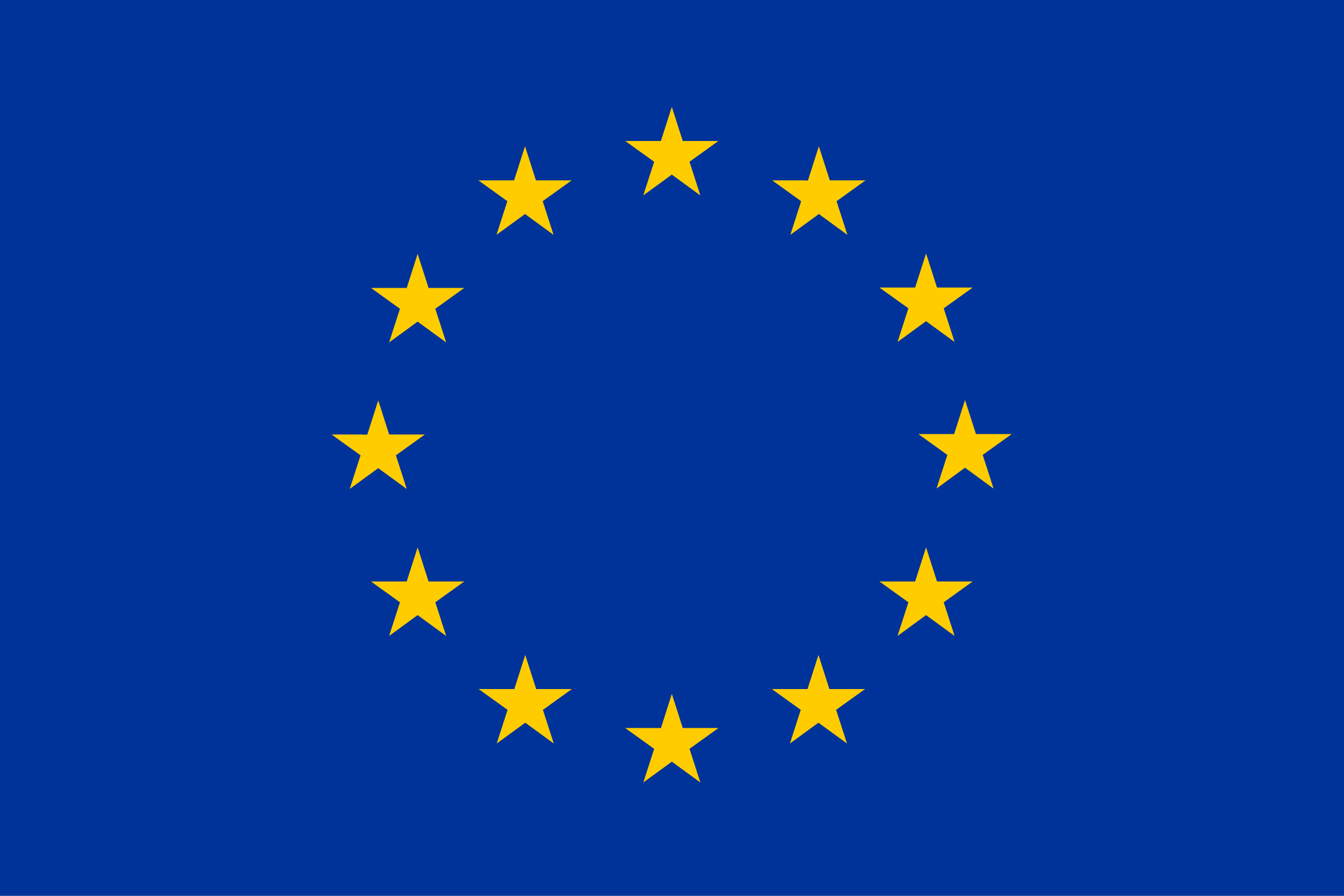}
}

\noindent This publication reflects only the views of the authors, the European Commission is not responsible for any use that may be made of the information it contains.

~\newline\noindent
MU has been supported by Quantum Valley Lower Saxony (QVLS).

\bmhead{Competing interests}
The authors have no competing interests to declare that are relevant to the content of this article.

\bmhead{Financial Interests}
The authors have no relevant financial or non-financial interests to disclose.

\bmhead{Ethics approval}
Ethical review and approval was not required for the study on human participants in accordance with the local legislation and institutional requirements.

\bibliography{sn-bibliography}


\begin{thebibliography}{21}
\ifx \bisbn   \undefined \def \bisbn  #1{ISBN #1}\fi
\ifx \binits  \undefined \def \binits#1{#1}\fi
\ifx \bauthor  \undefined \def \bauthor#1{#1}\fi
\ifx \batitle  \undefined \def \batitle#1{#1}\fi
\ifx \bjtitle  \undefined \def \bjtitle#1{#1}\fi
\ifx \bvolume  \undefined \def \bvolume#1{\textbf{#1}}\fi
\ifx \byear  \undefined \def \byear#1{#1}\fi
\ifx \bissue  \undefined \def \bissue#1{#1}\fi
\ifx \bfpage  \undefined \def \bfpage#1{#1}\fi
\ifx \blpage  \undefined \def \blpage #1{#1}\fi
\ifx \burl  \undefined \def \burl#1{\textsf{#1}}\fi
\ifx \doiurl  \undefined \def \doiurl#1{\url{https://doi.org/#1}}\fi
\ifx \betal  \undefined \def \betal{\textit{et al.}}\fi
\ifx \binstitute  \undefined \def \binstitute#1{#1}\fi
\ifx \binstitutionaled  \undefined \def \binstitutionaled#1{#1}\fi
\ifx \bctitle  \undefined \def \bctitle#1{#1}\fi
\ifx \beditor  \undefined \def \beditor#1{#1}\fi
\ifx \bpublisher  \undefined \def \bpublisher#1{#1}\fi
\ifx \bbtitle  \undefined \def \bbtitle#1{#1}\fi
\ifx \bedition  \undefined \def \bedition#1{#1}\fi
\ifx \bseriesno  \undefined \def \bseriesno#1{#1}\fi
\ifx \blocation  \undefined \def \blocation#1{#1}\fi
\ifx \bsertitle  \undefined \def \bsertitle#1{#1}\fi
\ifx \bsnm \undefined \def \bsnm#1{#1}\fi
\ifx \bsuffix \undefined \def \bsuffix#1{#1}\fi
\ifx \bparticle \undefined \def \bparticle#1{#1}\fi
\ifx \barticle \undefined \def \barticle#1{#1}\fi
\bibcommenthead
\ifx \bconfdate \undefined \def \bconfdate #1{#1}\fi
\ifx \botherref \undefined \def \botherref #1{#1}\fi
\ifx \url \undefined \def \url#1{\textsf{#1}}\fi
\ifx \bchapter \undefined \def \bchapter#1{#1}\fi
\ifx \bbook \undefined \def \bbook#1{#1}\fi
\ifx \bcomment \undefined \def \bcomment#1{#1}\fi
\ifx \oauthor \undefined \def \oauthor#1{#1}\fi
\ifx \citeauthoryear \undefined \def \citeauthoryear#1{#1}\fi
\ifx \endbibitem  \undefined \def \endbibitem {}\fi
\ifx \bconflocation  \undefined \def \bconflocation#1{#1}\fi
\ifx \arxivurl  \undefined \def \arxivurl#1{\textsf{#1}}\fi
\csname PreBibitemsHook\endcsname

\bibitem[\protect\citeauthoryear{Masiowski et~al.}{2022}]{masiowskiQuantumComputingFunding2022}
\begin{botherref}
\oauthor{\bsnm{Masiowski}, \binits{M.}},
\oauthor{\bsnm{Mohr}, \binits{N.}},
\oauthor{\bsnm{Soller}, \binits{H.}},
\oauthor{\bsnm{Zesko}, \binits{M.}}:
Quantum computing funding remains strong, but talent gap raises concern.
{{McKinsey Digital}}
(2022)
\end{botherref}
\endbibitem

\bibitem[\protect\citeauthoryear{{QUCATS}}{2024}]{qucatsQuantumFlagshipFuture2024}
\begin{botherref}
\oauthor{\bsnm{{QUCATS}}}:
Quantum {{Flagship}}: {{The}} Future Is {{Quantum}}.
https://qt.eu/
(2024)
\end{botherref}
\endbibitem

\bibitem[\protect\citeauthoryear{Greinert and M{\"u}ller}{2024}]{greinertEuropeanCompetenceFramework2024}
\begin{botherref}
\oauthor{\bsnm{Greinert}, \binits{F.}},
\oauthor{\bsnm{M{\"u}ller}, \binits{R.}}:
European {{Competence Framework}} for {{Quantum Technologies}}
(version 2.5)
(2024)
\doiurl{10.5281/zenodo.10976836}
\end{botherref}
\endbibitem

\bibitem[\protect\citeauthoryear{Greinert et~al.}{2023}]{greinertQuantumReadyWorkforce2023a}
\begin{botherref}
\oauthor{\bsnm{Greinert}, \binits{F.}},
\oauthor{\bsnm{M{\"u}ller}, \binits{R.}},
\oauthor{\bsnm{Goorney}, \binits{S.}},
\oauthor{\bsnm{Sherson}, \binits{J.}},
\oauthor{\bsnm{Ubben}, \binits{M.S.}}:
Towards a quantum ready workforce: The updated {{European Competence Framework}} for {{Quantum Technologies}}.
Frontiers in Quantum Science and Technology
\textbf{2}
(2023)
\doiurl{10.3389/frqst.2023.1225733}
\end{botherref}
\endbibitem

\bibitem[\protect\citeauthoryear{{Quantum Flagship}}{}]{quantumflagshipEuropeanQuantumReadiness}
\begin{botherref}
\oauthor{\bsnm{{Quantum Flagship}}}:
European {{Quantum Readiness Center}}.
https://quantumready.eu/\#/
\end{botherref}
\endbibitem

\bibitem[\protect\citeauthoryear{Sherson and Goorney}{2024}]{shersonDigiQDigitallyEnhanced2024}
\begin{botherref}
\oauthor{\bsnm{Sherson}, \binits{J.}},
\oauthor{\bsnm{Goorney}, \binits{S.}}:
{{DigiQ}}: {{Digitally Enhanced Quantum Technology Master}}.
https://www.digiq.eu/
(2024)
\end{botherref}
\endbibitem

\bibitem[\protect\citeauthoryear{{QURECA}}{2024}]{qurecaQTInduQuantumTechnologies2024}
\begin{botherref}
\oauthor{\bsnm{{QURECA}}}:
{{QTIndu}}: {{Quantum Technologies Courses}} for {{Industry}}.
https://qtindu.eu/
(2024)
\end{botherref}
\endbibitem

\bibitem[\protect\citeauthoryear{Kaur and {Venegas-Gomez}}{2022}]{kaurDefiningQuantumWorkforce2022}
\begin{barticle}
\bauthor{\bsnm{Kaur}, \binits{M.}},
\bauthor{\bsnm{{Venegas-Gomez}}, \binits{A.}}:
\batitle{Defining the quantum workforce landscape: A review of global quantum education initiatives}.
\bjtitle{Optical Engineering}
\bvolume{61}(\bissue{8}),
\bfpage{081806}
(\byear{2022})
\doiurl{10.1117/1.OE.61.8.081806}
\end{barticle}
\endbibitem

\bibitem[\protect\citeauthoryear{Fox et~al.}{2020}]{foxPreparingQuantumRevolution2020}
\begin{barticle}
\bauthor{\bsnm{Fox}, \binits{M.F.J.}},
\bauthor{\bsnm{Zwickl}, \binits{B.M.}},
\bauthor{\bsnm{Lewandowski}, \binits{H.J.}}:
\batitle{Preparing for the quantum revolution: {{What}} is the role of higher education?}
\bjtitle{Physical Review Physics Education Research}
\bvolume{16}(\bissue{2}),
\bfpage{020131}
(\byear{2020})
\doiurl{10.1103/PhysRevPhysEducRes.16.020131}
\end{barticle}
\endbibitem

\bibitem[\protect\citeauthoryear{Aiello et~al.}{2021}]{aielloAchievingQuantumSmart2021}
\begin{barticle}
\bauthor{\bsnm{Aiello}, \binits{C.D.}},
\bauthor{\bsnm{Awschalom}, \binits{D.D.}},
\bauthor{\bsnm{Bernien}, \binits{H.}},
\bauthor{\bsnm{Brower}, \binits{T.}},
\bauthor{\bsnm{Brown}, \binits{K.R.}},
\bauthor{\bsnm{Brun}, \binits{T.A.}},
\bauthor{\bsnm{Caram}, \binits{J.R.}},
\bauthor{\bsnm{Chitambar}, \binits{E.}},
\bauthor{\bsnm{Felice}, \binits{R.D.}},
\bauthor{\bsnm{Edmonds}, \binits{K.M.}},
\bauthor{\bsnm{Fox}, \binits{M.F.J.}},
\bauthor{\bsnm{Haas}, \binits{S.}},
\bauthor{\bsnm{Holleitner}, \binits{A.W.}},
\bauthor{\bsnm{Hudson}, \binits{E.R.}},
\bauthor{\bsnm{Hunt}, \binits{J.H.}},
\bauthor{\bsnm{Joynt}, \binits{R.}},
\bauthor{\bsnm{Koziol}, \binits{S.}},
\bauthor{\bsnm{Larsen}, \binits{M.}},
\bauthor{\bsnm{Lewandowski}, \binits{H.J.}},
\bauthor{\bsnm{McClure}, \binits{D.T.}},
\bauthor{\bsnm{Palsberg}, \binits{J.}},
\bauthor{\bsnm{Passante}, \binits{G.}},
\bauthor{\bsnm{Pudenz}, \binits{K.L.}},
\bauthor{\bsnm{Richardson}, \binits{C.J.K.}},
\bauthor{\bsnm{Rosenberg}, \binits{J.L.}},
\bauthor{\bsnm{Ross}, \binits{R.S.}},
\bauthor{\bsnm{Saffman}, \binits{M.}},
\bauthor{\bsnm{Singh}, \binits{M.}},
\bauthor{\bsnm{Steuerman}, \binits{D.W.}},
\bauthor{\bsnm{Stark}, \binits{C.}},
\bauthor{\bsnm{Thijssen}, \binits{J.}},
\bauthor{\bsnm{Vamivakas}, \binits{A.N.}},
\bauthor{\bsnm{Whitfield}, \binits{J.D.}},
\bauthor{\bsnm{Zwickl}, \binits{B.M.}}:
\batitle{Achieving a quantum smart workforce}.
\bjtitle{Quantum Science and Technology}
\bvolume{6}(\bissue{3}),
\bfpage{030501}
(\byear{2021})
\doiurl{10.1088/2058-9565/abfa64}
\end{barticle}
\endbibitem

\bibitem[\protect\citeauthoryear{Hughes et~al.}{2022}]{hughesAssessingNeedsQuantum2022}
\begin{barticle}
\bauthor{\bsnm{Hughes}, \binits{C.}},
\bauthor{\bsnm{Finke}, \binits{D.}},
\bauthor{\bsnm{German}, \binits{D.-A.}},
\bauthor{\bsnm{Merzbacher}, \binits{C.}},
\bauthor{\bsnm{Vora}, \binits{P.M.}},
\bauthor{\bsnm{Lewandowski}, \binits{H.J.}}:
\batitle{Assessing the {{Needs}} of the {{Quantum Industry}}}.
\bjtitle{IEEE Transactions on Education}
\bvolume{65}(\bissue{4}),
\bfpage{592}--\blpage{601}
(\byear{2022})
\doiurl{10.1109/TE.2022.3153841}
\end{barticle}
\endbibitem

\bibitem[\protect\citeauthoryear{Hasanovic et~al.}{2022}]{hasanovicQuantumTechnicianSkills2022}
\begin{barticle}
\bauthor{\bsnm{Hasanovic}, \binits{M.}},
\bauthor{\bsnm{Panayiotou}, \binits{C.A.}},
\bauthor{\bsnm{Silberman}, \binits{D.M.}},
\bauthor{\bsnm{Stimers}, \binits{P.}},
\bauthor{\bsnm{Merzbacher}, \binits{C.I.}}:
\batitle{Quantum technician skills and competencies for the emerging {{Quantum}} 2.0 industry}.
\bjtitle{Optical Engineering}
\bvolume{61}(\bissue{8}),
\bfpage{081803}
(\byear{2022})
\doiurl{10.1117/1.OE.61.8.081803}
\end{barticle}
\endbibitem

\bibitem[\protect\citeauthoryear{Greinert et~al.}{2023}]{greinertFutureQuantumWorkforce2023}
\begin{barticle}
\bauthor{\bsnm{Greinert}, \binits{F.}},
\bauthor{\bsnm{M{\"u}ller}, \binits{R.}},
\bauthor{\bsnm{Bitzenbauer}, \binits{P.}},
\bauthor{\bsnm{Ubben}, \binits{M.S.}},
\bauthor{\bsnm{Weber}, \binits{K.-A.}}:
\batitle{Future quantum workforce: {{Competences}}, requirements, and forecasts}.
\bjtitle{Physical Review Physics Education Research}
\bvolume{19}(\bissue{1}),
\bfpage{010137}
(\byear{2023})
\doiurl{10.1103/PhysRevPhysEducRes.19.010137}
\end{barticle}
\endbibitem

\bibitem[\protect\citeauthoryear{Cameron}{2009}]{cameronSequentialMixedModel2009}
\begin{barticle}
\bauthor{\bsnm{Cameron}, \binits{R.}}:
\batitle{A sequential mixed model research design: {{Design}}, analytical and display issues}.
\bjtitle{International Journal of Multiple Research Approaches}
\bvolume{3}(\bissue{2}),
\bfpage{140}--\blpage{152}
(\byear{2009})
\doiurl{10.5172/mra.3.2.140}
\end{barticle}
\endbibitem

\bibitem[\protect\citeauthoryear{Greinert}{2024}]{greinertSupplementaryMaterialQuantum2024}
\begin{botherref}
\oauthor{\bsnm{Greinert}, \binits{F.}}:
Supplementary {{Material}} for ``Advancing quantum technology workforce: industry insights into qualification and training needs''.
Zenodo
(2024).
\doiurl{10.5281/zenodo.12531594}
\end{botherref}
\endbibitem

\bibitem[\protect\citeauthoryear{Mayring}{2014}]{mayringQualitativeContentAnalysis2014}
\begin{botherref}
\oauthor{\bsnm{Mayring}, \binits{P.}}:
Qualitative {{Content Analysis}}: Theoretical Foundation, Basic Procedures and Software Solution,
Klagenfurt
(2014)
\end{botherref}
\endbibitem

\bibitem[\protect\citeauthoryear{R{\"a}diker and Kuckartz}{2020}]{radikerFocusedAnalysisQualitative2020}
\begin{bbook}
\bauthor{\bsnm{R{\"a}diker}, \binits{S.}},
\bauthor{\bsnm{Kuckartz}, \binits{U.}}:
\bbtitle{Focused {{Analysis}} of {{Qualitative Interviews}} With {{MAXQDA}}},
\bedition{1}st edn.
\bpublisher{MAXQDA Press},
\blocation{Berlin}
(\byear{2020})
\end{bbook}
\endbibitem

\bibitem[\protect\citeauthoryear{Landis and Koch}{1977}]{landisMeasurementObserverAgreement1977}
\begin{barticle}
\bauthor{\bsnm{Landis}, \binits{J.R.}},
\bauthor{\bsnm{Koch}, \binits{G.G.}}:
\batitle{The {{Measurement}} of {{Observer Agreement}} for {{Categorical Data}}}.
\bjtitle{Biometrics}
\bvolume{33}(\bissue{1}),
\bfpage{159}--\blpage{174}
(\byear{1977})
\doiurl{10.2307/2529310}
\end{barticle}
\endbibitem

\bibitem[\protect\citeauthoryear{Robbins and Heiberger}{2011}]{robbinsPlottingLikertOther2011}
\begin{botherref}
\oauthor{\bsnm{Robbins}, \binits{N.B.}},
\oauthor{\bsnm{Heiberger}, \binits{R.M.}}:
Plotting {{Likert}} and {{Other Rating Scales}}.
Proceedings of the 2011 Joint Statistical Meeting,
1058--1066
(2011)
\end{botherref}
\endbibitem

\bibitem[\protect\citeauthoryear{{QuIC}}{2024}]{quicEuropeanQuantumIndustry2024}
\begin{botherref}
\oauthor{\bsnm{{QuIC}}}:
European {{Quantum Industry Consortium}}.
https://www.euroquic.org/
(2024)
\end{botherref}
\endbibitem

\bibitem[\protect\citeauthoryear{Henke et~al.}{2018}]{henkeAnalyticsTranslatorNew2018}
\begin{botherref}
\oauthor{\bsnm{Henke}, \binits{N.}},
\oauthor{\bsnm{Levine}, \binits{J.}},
\oauthor{\bsnm{McInerney}, \binits{P.}}:
Analytics translator: {{The}} new must-have role.
{{McKinsey Analytics}}
(2018)
\end{botherref}
\endbibitem

\end{thebibliography}

\newpage
\begin{appendices}
\small
\section{Additions on the survey instruments}
\subsection{Materials to show in the interviews}\label{app:interwMaterials}
\begin{figure}[ht]
    \centering
    \includegraphics[width=0.7\textwidth]{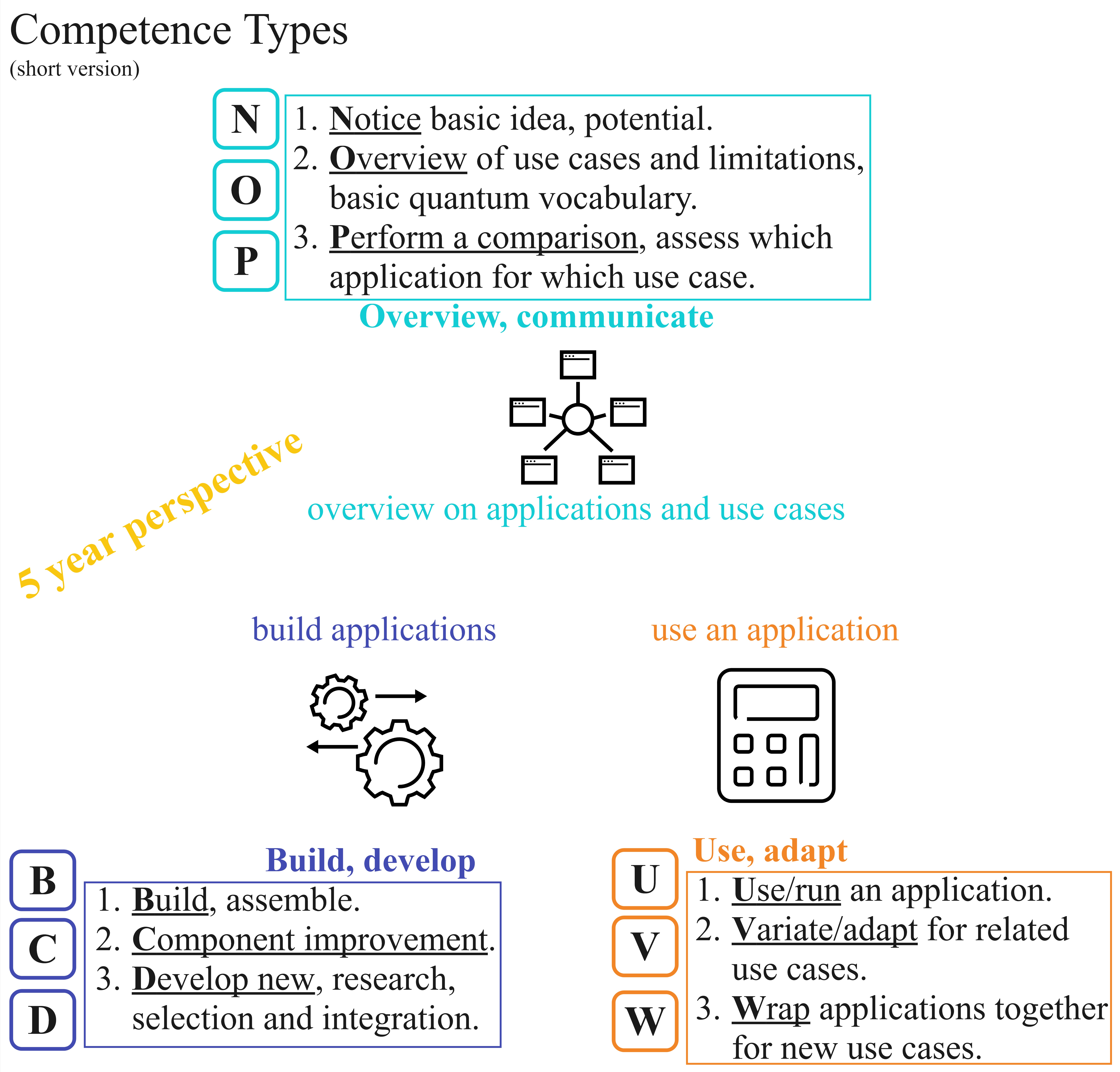}
    \caption{Competence types: short version from July 2023. Outdated.}
    \label{fig:CompTypes}
    \vspace{-0.5cm}
\end{figure}

\begin{figure}[ht]
    \centering
    \includegraphics[width=0.7\textwidth]{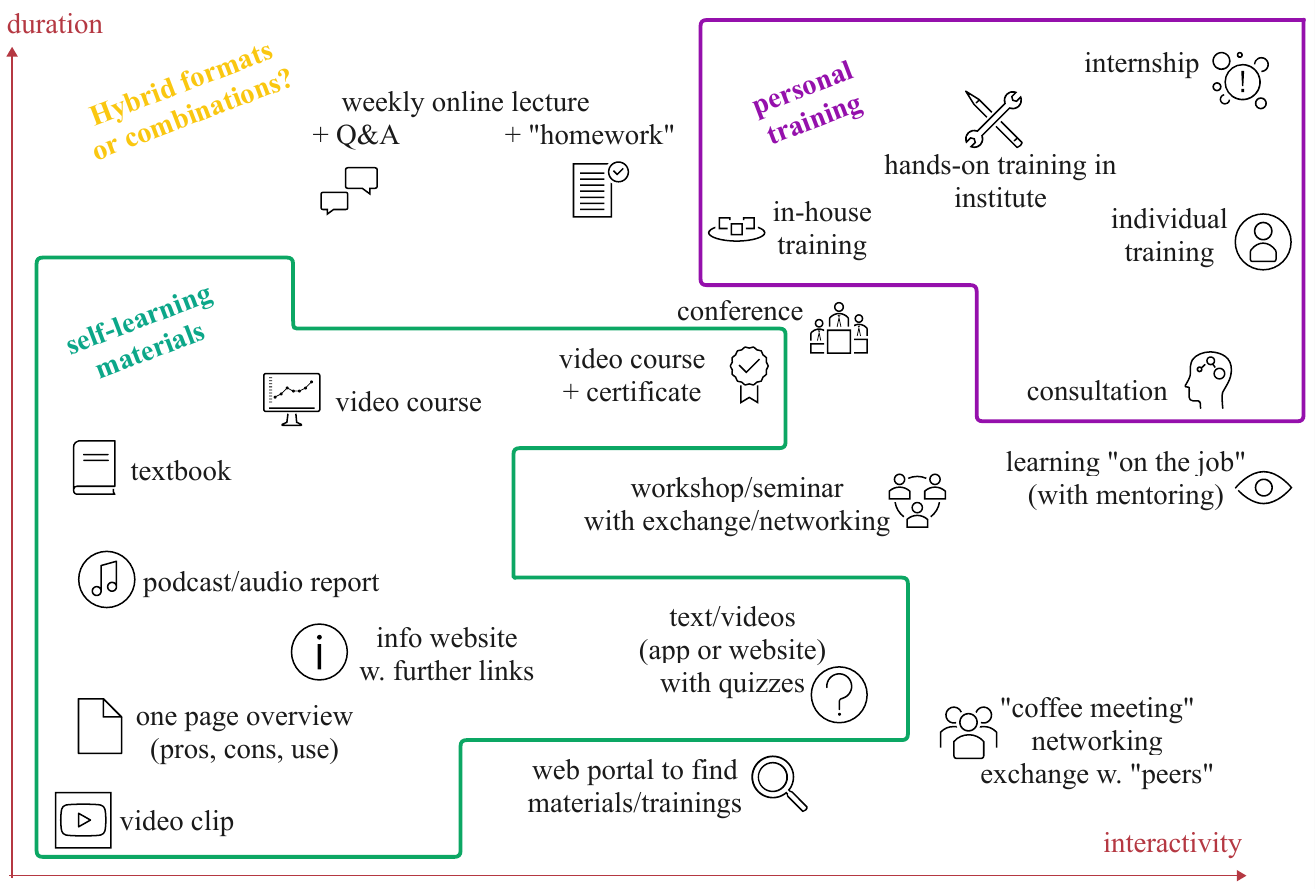}
    \caption{Training formats: map of training formats provided in some interviews as inspiration to discuss preferred formats.}
    \label{fig:TrainForm}
\end{figure}

Figures~\ref{fig:CompTypes} and \ref{fig:TrainForm} were prepared to support the interviews, optionally to be shown depending on the interview flow. Fig.~\ref{fig:CompTypes} is a shortened version, in the interviews, longer descriptions and examples for each level were available. These competence types turned out to be an interim step towards the Qualification Profiles that were added to the \CF\ within the update to version 2.5, so they will not be used any further.

\subsection{Rating items in the follow-up questionnaire}\label{subsec:items}
In the follow-up questionnaire, 19 items were presented to be assessed as  `high need', `low need', or `no need' \textbf{--} or `can´t assess'. Tab.~\ref{tab:ratings} summarizes the ratings statistics. The option `can't assess' includes the up to two counts for individuals who did not respond to that particular item. The ratings were clustered in terms of targeting different interest groups: 

\begin{enumerate}
    \item For business people and colleagues of quantum experts ...
    \begin{enumerate}
        \item ... to ``learn the quantum language'', without details of math and physics, to develop a basic understanding of what the quantum people are talking about and to enable efficient communication.
        \item ... to understand the ``properties'' of quantum, i.e. fundamental phenomena such as superposition and entanglement.
        \item ... to have an idea of the quantum technology landscape: what technologies are being developed, what are the (expected) applications and use cases, and why are they or could they become relevant to their own business.
        \item ... to develop a critical perspective on the quantum hype, a realistic picture of (im)possibilities and timelines.
        \item ... to have training in the native language, as this makes it easier to understand the principles, rather than having everything in English.
        \item ... to obtain a certificate at this introductory level.
        \item ... to have face-to-face training to network and talk with peers, to see that others have similar problems, and to ask ``stupid'' questions.
    \end{enumerate}
    \begin{samepage}
    \item For decision makers, politicians, etc. ...
    \begin{enumerate}
        \item ... to get easily accessible information about what they (might) be able to do with quantum technologies, what value they (might) bring, and why they should care. Any math or physics is inconvenient and to be avoided.
        \item ... to have it in their native language to facilitate understanding and increase trust.
    \end{enumerate}
    \end{samepage}
    \item For engineers to go deeper ...
    \begin{enumerate}
        \item ... to have specific training that is truly tailored to the trainees.
        \item ... to have lecture-like training or materials on quantum theory, with details on the physics behind quantum technologies and advanced mathematics.
        \item ... to get hands-on training in working with quantum hardware, to gain lab experience, such as setting up measurements with an NV center.
        \item ... to be trained in how to use a quantum device and integrate it into their work.
        \item ... to receive a certificate after attending a course in which a specific skill is trained.
        \item ... to have everything in English to communicate internationally.
    \end{enumerate}
    \item Provide online resources like:
    \begin{enumerate}
        \item A curated selection of good online material on quantum technologies and quantum theory (YouTube videos, podcasts, blogs, …).
        \item A brief overview of quantum technology approaches and available solutions with comparison (pros \& cons).
    \end{enumerate}
    \item Get more traditional engineers interested in working in the quantum domain.
    \item Get more school students interested in STEM and quantum. 
\end{enumerate}
\begin{table}[ht]
    \centering
    \caption{Rating statistics for the 55 responses on the follow-up survey.}
    \label{tab:ratings}
    \begin{tabular}{c|c|c|c|c}
        &high &low &no &can't \\
        item (short)& need & need& need& assess\\
        \hline
        1 (a) business: learn language&41&9&2&3\\
        1 (b) business: understand properties&19&30&5&1\\
        1 (c) business: technical landscape&46&9&0&0\\
        1 (d) business: critical perspective (hype)&41&11&2&1\\
        1 (e) native language for business/colleagues&9&19&21&6\\
        1 (f) certificate for business (introductory level)&8&25&18&4\\
        1 (g) business: face-to-face training&21&24&4&6\\
        \hline
        2 (a) decision makers: value&43&7&1&4\\
        2 (b) native language for decision makers&19&15&12&9\\
        \hline
        3 (a) engineers: tailored training&39&11&0&5\\
        3 (b) engineers: q. theory (lecture-like)&19&28&4&4\\
        3 (c) engineers: hands-on training&31&15&4&5\\
        3 (d) engineers: how to use&31&17&2&5\\
        3 (e) certificate for engineers (specific skill)&14&23&10&8\\
        3 (f) English for engineers&26&13&6&10\\
        \hline
        4 (a) curated selection&38&15&1&1\\
        4 (b) overview, comparison&42&11&1&1\\
        \hline
        5 traditional engineers&36&16&0&3\\
        \hline
        6 school students in STEM&40&11&1&3\\
    \end{tabular}
\end{table}

\section{Category system}\label{app:categories}
This section provides the category system for the qualitative content analysis of the interview transcripts. It is reduced to those codes that are relevant to the results discussed in this paper. Other codes documenting, for example, the company type or country, are not shown; these numbers are given in Sec.~\ref{subsubsec:interviewParticipants}. The subcategories and quantitative data for the discussed jobs (1.2) are already shown in Tab~\ref{tab:professions}. 

For each code, the [number] indicates how many interview segments were coded with that code. No number means that this is a supporting code, which only helps to structure the codes on the next level. Therefore, they were not scored for intercoder agreement.

\renewcommand{\labelenumii}{\arabic{enumi}.\arabic{enumii}}
\renewcommand{\labelenumiii}{\arabic{enumi}.\arabic{enumii}.\arabic{enumiii}}
\renewcommand{\labelenumiv}{\arabic{enumi}.\arabic{enumii}.\arabic{enumiii}.\arabic{enumiv}}
\begin{enumerate}
    \item workforce needs 
    \begin{enumerate}
        \item workforce development
        \begin{enumerate}
            \item challenges in workforce development
            \begin{enumerate}
                \item start-up to scale-up change \dotfill [11]
                \item challenges in attracting/finding talent \dotfill [13]
            \end{enumerate}
            \item strategies for getting skilled people \dotfill [18]
            \item hiring and/or upskilling? \dotfill [11]
            \item on (required) degrees \dotfill [18]
            \item other challenges in workforce development \dotfill [23]
        \end{enumerate}
        \item jobs/disciplines in own company \dotfill see Tab~\ref{tab:professions}
    \end{enumerate}
    \item roles/profiles with training needs 
    \footnote{These categories were used to formulate a first version of Qualification Profiles to extend the \CF, see Sec.~\ref{subsec:qualiProf}.}
    \begin{enumerate}
        \item general or content-focused remarks, challenges \dotfill [6]
        \item quantum aware workforce
        \begin{enumerate}
            \item basic user (click a button) \dotfill [6]
            \item engineers with (almost) no quantum \dotfill [22]
            \item hype, basic idea (not) for all, incl. admin people \dotfill [24]
        \end{enumerate}
        \item quantum literate workforce
        \begin{enumerate}
            \item advanced user (with adaption) \dotfill [23]
            \item management (non-QT comp.), business, sales, policy makers \dotfill [42]
            \item engineers with overview on QT (also e.g. cryo physicists) \dotfill [65]
            \begin{itemize}
                \item programming \dotfill [8]
            \end{itemize}
            \item use-case identifier \dotfill [19]
        \end{enumerate}
        \item quantum expert workforce (needs QT study program or PhD)
        \begin{enumerate}
            \item strategist, consultants (comparison, assessment) \dotfill [27]
            \item customer/sales engineers \dotfill [12]
            \item systems engineers (integration, industrialisation) \dotfill [13]
            \item quantum experts / expertise needs \dotfill [35]
            \item quantum algorithm developers (high-level) \dotfill [22]
        \end{enumerate}
    \end{enumerate}
    \item training format needs
    \begin{enumerate}
        \item training suggestions or how it is done in the company
        \begin{enumerate}
            \item internal \dotfill [30]
            \item hands-on training, workshop \dotfill [14]
            \item conference or other in-person events, focus on networking \dotfill [7]
            \item university courses/seminars  \dotfill [15]
            \item online course or similar \dotfill [17]
            \item other/multi \dotfill [10]
        \end{enumerate}
        \item boundary conditions for (external) training
        \begin{enumerate}
            \item language
            \begin{enumerate}
                \item English \dotfill [18]
                \item local language \dotfill [2]
                \item undetermined \dotfill [6]
            \end{enumerate}
            \item certificate \dotfill [22]
            \item costs \dotfill [11]
            \item duration/time splitting \dotfill [11]
        \end{enumerate}
        \item on self-learning
        \begin{enumerate}
            \item self-learning materials: preferred formats
            \begin{enumerate}
                \item multi \dotfill [7]
                \item podcast \dotfill [10]
                \item written text \dotfill [4]
                \begin{itemize}
                    \item NOT read [3]; \hfill info webpage (wiki) [2]; \hfill one page overview [4];\newline book, script \dotfill [5]; \hfill scientific documents, reports, etc. \dotfill [3]
                \end{itemize}
                \item videos \dotfill [8]
                \begin{itemize}
                    \item short video clips \dotfill [4]; \hfill video course \dotfill [5]
                \end{itemize}
            \end{enumerate}
            \item interactive elements \dotfill [6]
            \item support finding materials \dotfill [11]
        \end{enumerate}
    \end{enumerate}
    \item additions and advice
    \begin{enumerate}
        \item non-quantum needs \dotfill [4]
        \begin{enumerate}
            \item needs of professional skills, e.g. problem solving \dotfill [3]
            \begin{itemize}
                \item communication skills \dotfill [7]
            \end{itemize}
            \item needs in business \dotfill [18]
            \item needs in other STEM \dotfill [12]
        \end{enumerate}
        \item advice
        \begin{enumerate}
            \item talk to industry/'user' of training \dotfill [10]
            \item other suggestions (on training design) \dotfill [27]
        \end{enumerate}
    \end{enumerate}
\end{enumerate}

\end{appendices}
\clearpage


\end{document}